\documentclass[reprint,onecolumn, showpacs,preprintnumbers,amsmath,amssymb,nofootinbib]{revtex4-2} 
\usepackage{subfiles}
\usepackage{amsmath}
\usepackage{amssymb}
\usepackage{amsfonts}
\usepackage[caption=false,position=top]{subfig}
\usepackage{graphicx}
\usepackage{color}
\usepackage{float}
\usepackage{comment}
\usepackage[normalem]{ulem}
\usepackage{ascmac}
\usepackage{framed}
\usepackage{multirow}
\usepackage{mathtools}
\usepackage{bm}
\usepackage{array}
\usepackage[%
 setpagesize=false,%
 bookmarks=true,%
 bookmarksdepth=tocdepth,%
 bookmarksnumbered=true,%
 colorlinks=true,%
 linkcolor=red,%
 citecolor=DarkBlue,%
 linktocpage =true,%
 pdftitle={},%
 pdfsubject={},%
 pdfauthor={},%
 pdfkeywords={}%
]{hyperref}
\graphicspath{{figs/}}

\definecolor{DarkBlue}{rgb}{0,0,0.7}

\newcommand{\figref}[1]{Fig.\ref{#1}}
\newcommand{\figsref}[1]{Figs.\ref{#1}}

\renewcommand{\Re}{\textrm{Re}}
\renewcommand{\Im}{\textrm{Im}}
\newcommand{\VN}{{\cal V_N}}
\newcommand{\VS}{{\cal V_S}}
\newcommand{\E}{\boldsymbol{E}}
\newcommand{\D}{\boldsymbol{D}}
\newcommand{\B}{\boldsymbol{B}}

\begin{document}
\title{\Large \bf Vacuum Magnetospheres around Kerr Black Holes with a Thin Disk}
\author{\vspace{1cm}
\Large Yota Endo}
\email{d20sa003@st.osaka-cu.ac.jp}
\affiliation{
Department of Mathematics and Physics,
Graduate School of Science, Osaka Metropolitan University, Osaka 558-8585, Japan}
\author{\Large Hideki Ishihara}
\email{h.ishihara@omu.ac.jp}
\affiliation{
Nambu Yoichiro Institute of Theoretical and Experimental Physics (NITEP),
Osaka Central Advanced Mathematical Institute,
Osaka Metropolitan University, Osaka 558-8585, Japan}
\author{\Large Masaaki Takahashi}
\email{mtakahas@auecc.aichi-edu.ac.jp}
\affiliation{Department of Physics and Astronomy, Aichi University of Education, Kariya, Aichi 448-8542, Japan}
\hfill{OCU-PHYS-607}

\hfill{AP-GR-203}

\hfill{NITEP 234}

\begin{abstract}
  We construct a model for stationary and axisymmetric black hole magnetospheres by solving the vacuum Maxwell equations in Kerr backgrounds. The poloidal magnetic field is generated by a toroidal electric current in a thin disk on the equatorial plane with an inner edge, and the poloidal electric field is induced due to the black hole spin.
  We assume a current distribution whose direction reverses at a radius on the disk. 
  The magnetospheric structure is divided into the inner region and the outer one due to the current reversal. 
  In the inner region, some magnetic field lines connect the black hole and the disk, and others surround the inner edge of the disk. In the outer region, the field lines connect the disk and infinity. 
  We investigate the black hole spin dependence of the magnetospheric structure. 
  In the magnetosphere around the Kerr black hole, the electric fields are generated by the black hole spin. 
  In our model, the charge distribution is induced on the disk and the event horizon. Except near the black hole, the electric field strength is proportional to the black hole spin, while the direction of the electric field remains almost the same regardless of the spin. 
  On the other hand, the electromagnetic field near the event horizon strongly depends on the spin. 
  In the maximally rotating case, the magnetic field lines are expelled from the event horizon and the electric equipotential surface coincides with the horizon. 
\end{abstract}

\maketitle
\pagebreak

\section{Introduction}
The global structure of electromagnetic fields surrounding a black hole is important in high-energy astrophysical objects such as active galactic nuclei (AGNs), compact X-ray sources, and gamma-ray bursts (see \cite{Punsly:2008hj,Camenzind:2007sa} for reviews). 
Interactions of magnetic plasma in strong gravitational and electromagnetic fields would be directly related to many astrophysical phenomena, that is, the generation of high-energy radiations and cosmic rays, the collimation and acceleration of outgoing jets and winds, and so on.  


We expect that a region where the energy density of magnetic field dominates the energy density of plasma should be created around a black hole with an accretion disk. Such a region is called a black hole magnetosphere.
Rotational energy extraction mechanisms from the black hole magnetosphere are often proposed as an energy source for high-energy phenomena. 
In particular, attention is focused on the Blandford-Znajek process \cite{Blandford:1977ds} for relativistic jets and the magnetized Penrose process \cite{Wagh:1989zqa,Tursunov:2020juz} for high-energy cosmic rays. 
The energy extraction efficiency by these processes in the black hole magnetosphere depends largely on the strength and configuration of the magnetic field (e.g., \cite{Phinney:1983rg,Thoelecke:2017jrz,Thoelecke:2019tqg,Kolos:2020gdc}), in addition to the black hole's spin.   
Numerous studies (e.g., \cite{Blandford:1976gf,Beskin:1997ga,Camenzind:1987sf,Mobarry:1986tl}) have been conducted to clarify the structures of the electromagnetic field in black hole magnetospheres. 

For modeling the black hole magnetosphere, general relativistic magnetohydrodynamics (GR-MHD) has been applied. 
Since the basic equations of GR-MHD system involve highly nonlinear second partial differential equations (see, e.g., \cite{Takahashi:1990hk,Nitta:1991ui}), it is hard to solve them analytically. 
Therefore, numerical methods are used to solve the global behavior of the GR-MHD system \cite{Koide:2003fj,McKinney:2004ka,McKinney:2006tf} (see also \cite{Beskin:2009hd,Davis:2020wea} for a review).  
However, the setting of initial conditions and the resulting final state (e.g. quasi-steady state) are complicated and debatable, and it is difficult to understand the effects of plasma inertia, electromagnetic fields and curved spacetime, separately.
In this paper, we concentrate on the interaction between the gravitational field and the electromagnetic field, and solve the structure of electromagnetic fields in stationary and axisymmetric magnetospheres around rotating black holes. 
Here, we assume that no plasma exists within the magnetosphere except for current sources on the equatorial plane (that is, a vacuum black hole magnetosphere). 
In particular, we clarify how the spacetime dragging effect affects the structure of the electromagnetic field near the black hole. 

The vacuum magnetosphere solution in a Kerr spacetime created by a ring current placed on the equatorial plane is investigated analytically by many authors \cite{Chitre:1975sc,Petterson:1975sg,Znajek:1978ht,Bicak:1976ag,Moss:2011tc}. 
The basic equations are derived by using the Newman and Penrose formalism \cite{Newman:1961qr,Teukolsky:1973ha}, where the variables are separated into radial and polar directions, and the solution is described by a superposition of the Legendre polynomials \cite{Petterson:1975sg}. 
In the solution, the regularity on the rotation axis of the black hole is assumed as the boundary condition for the polar angle. 
On the other hand, in the radial direction, the Legendre polynomials are singular either at the event horizon or at spatial infinity. 
Therefore, the solutions should be described in two subsets of the Legendre functions; 
the ``inner functions'', which are regular at the event horizon but singular at infinity, and the ``outer functions'', which are regular at infinity but singular at the event horizon.  The inner solution, expanded by ``inner functions'' and the outer solution, expanded by ``outer functions'' are matched at the spherical surface with the radius of the ring current. In terms of the infinite sum of ring currents on the equatorial plane with various radii, Li \cite{Li:1999ke} discussed a black hole magnetosphere with a disk current. 

Tomimatsu and Takahashi \cite{Tomimatsu:2000yp} (hereafter TT01) discussed a magnetosphere created by toroidal currents flowing on a geometrically thin disk placed on the equatorial plane in a Schwarzschild spacetime. 
In TT01, solutions of the Maxwell equation are expanded in a set of the Legendre functions with complex indices that are regular both at the event horizon and at infinity but are singular on the rotation axis in one of the southern or northern hemispheres. 
In TT01, the solutions are divided into ``northern'' and ``southern'' solutions; the former is expanded by ``north functions'' that are regular in the northern region but singular at the southern part of the rotation axis, and the latter is expanded by ``south functions'' that are regular in the southern region but singular at the northern part of the rotation axis. 
To construct a whole solution, the northern and southern solutions are matched on the equatorial plane by using a junction condition given by the current distribution on the equatorial plane.  

In TT01, it is assumed that the thin disk has an inner edge, i.e., no current exists between the inner edge of the disk and the event horizon. 
In addition, the direction of the current is reversed at a certain radius on the disk. 
Therefore, the magnetic field generated by the disk current is divided into two parts by an almost spherical separatrix surface. 
In the inner region, magnetic field lines are confined inside the separatrix surface; loop-shaped magnetic field lines form around the inner edge of the disk, and some magnetic field lines connect the disk surface and the event horizon. 
On the other hand, in the outer region of the separatrix surface, the magnetic field lines extend from the disk surface, and take on a conical shape at a sufficiently far distance. 

In this paper, extending the work in TT01, we obtain the electromagnetic field for vacuum black hole magnetospheres with a disk current in Kerr spacetimes by the use of the Newman-Penrose formalism to separate variables in the Maxwell equations. 
The functions that expand the solutions differ from those used in the case of the ring current. 
In order to clarify how the black hole spin affects the structure of the electromagnetic field, we fix the radial distribution of the toroidal current and the radius of the inner edge. 

Even in the Kerr geometry, the global structure of the magnetic field is shown to be similar to that in the Schwarzschild geometry in the region with a radius larger than the inner edge radius, regardless of the spin parameter. 
The radius of the separatrix surface and the shape of magnetic field lines at a distant region do not depend on the black hole spin. 
On the other hand, the structure of magnetic fields near the black hole is almost parallel to the rotation axis for a slowly rotating black hole case, while for the extreme Kerr black hole case magnetic field lines are expelled from the black hole; this phenomenon is known as the ``black hole Meissner effect'' \cite{Bicak:1985af,Komissarov:2007rc}.
The structure near the black hole is almost the same as that in the Wald solution \cite{Wald:1974np} in which there is no source everywhere (see appendix \ref{app:wald} for brief review). The similar structure appears in the ring current case, in which there is no source near the black hole.

The most important difference between the magnetosphere in the Kerr spacetime and that in the Schwarzschild \cite{Tomimatsu:2000yp} is that an electric field is generated in the magnetosphere due to the rotation of the black hole. 
The amplitude of the electric field is proportional to the black hole spin and the direction of the electric field does not depend on the black hole spin, except near the event horizon. 
We see that the electric potential has a saddle point on the rotation axis inside the separatrix surface of the magnetic field.
We also see that the electric flux density is generated in the vacuum magnetosphere as if charge separation occurs on the surfaces of the black hole and on the thin disk. 
Furthermore, in our model, when the spin parameter takes the extreme value, the event horizon becomes an equipotential surface of the electric field in addition to the occurrence of the black hole Meissner effect.

\textbf{In Sec.\ref{sec:Maxeq}}, we derive the basic equations for a stationary and axisymmetric vacuum electromagnetic field 
in Kerr spacetimes with a thin disk current on the equatorial plane by using the Newman-Penrose formalism for Maxwell's field equations.  
\textbf{In Sec.\ref{sec:bc}}, the equations are solved analytically, and the northern and southern vacuum solutions are matched at the equatorial plane according to the junction condition. 
\textbf{In Sec.\ref{sec:num_any}}, we show some configurations of the magnetic and electric fields by numerical plots. We consider the possibility of the acceleration of charged particles and the accretion of charged particles from the disk to the black hole. 
Finally,  \textbf{in Sec.\ref{sec:dscss}}, we summarize our conclusions. We also discuss the possible extension of the vacuum magnetosphere to force-free magnetospheres.

\section{The Maxwell equations in the Newman Penrose formalism}
\label{sec:Maxeq}
\subsection{Basic equations}
We consider a stationary axisymmetric vacuum black hole magnetosphere in the Kerr spacetimes. 
The metric in Boyer-Lindquist coordinates $(t,r,\theta,\varphi)$ with $c=G=4\pi\epsilon_0=1$ is given by 
\begin{align}
    ds^2 = &-\left(1-\frac{2Mr}{\Sigma}\right)dt^2
	-\frac{4Mra\sin^2\theta}{\Sigma}dtd\varphi\notag\\
    & \qquad+\frac{\Sigma}{\Delta}dr^2+\Sigma d\theta^2
	+\left(r^2+a^2+\frac{2Mr}{\Sigma}a^2\sin^2\theta\right)\sin^2\theta d\varphi^2,
\label{met:Kerr}
\end{align}
where $M$ is the mass of the black hole, $a$ is the spin parameter, and 
\begin{align*}
    &\Sigma \coloneqq r^2+a^2\cos^2\theta,
\\ 
	&\Delta \coloneqq r^2-2Mr+a^2 =(r-r_+)(r-r_-),
\\
 	&r_\pm \coloneqq M\pm\sqrt{M^2-a^2} .
\end{align*}
The event horizon radius, $r_\mathrm{H}$, is given by $r_\mathrm{H}=r_+$. 

The Maxwell equations we consider in this paper are 
\begin{align}
	&\frac{1}{\sqrt{-g}}\partial_\mu\left(\sqrt{-g} F^{\mu\nu}\right) 
		= -4\pi j^\nu \delta(\theta-\pi/2),
\label{eq:Maxeq}
\\
	&\epsilon^{\sigma\mu\nu\lambda}\partial_\mu F_{\nu\lambda}=0,
\label{eq:Maxid}
\end{align}
where the field strength tensor of a vector potential $A_\mu$ is defined by 
$F_{\mu\nu} \coloneqq \partial_\mu A_\nu-\partial_\nu A_\mu$,  
$j^\mu$ is a current surface density on the equatorial plane, $\theta=\pi/2$, $\epsilon_{\alpha\beta\gamma\delta}$ is the Levi-Civita tensor, $\epsilon_{0123} =\sqrt{-g}$,
and $g = \det g_{\mu\nu}$. 
It is assumed that $F_{\mu\nu}$ depends on $r$ and $\theta$, and 
$j^\nu$ depends on $r$.

We obtain solutions for the two vacuum bulk regions: 
the region above the equatorial plane, $0<\theta<\pi/2$, 
and the region below the plane, $\pi/2<\theta<\pi$. 
We refer to these as the northern region, $\VN$, and the southern region, $\VS$, respectively.  
These two vacuum solutions are matched at the equatorial plane, $\theta=\pi/2$, 
so that the total field follows the Maxwell equations with currents on the thin disk.

By the assumption that the current $j^\nu$ resides on the thin disk at the equatorial plane, 
the integrations of \eqref{eq:Maxeq} and \eqref{eq:Maxid} by $\theta$ in the tiny interval $[\pi/2-\epsilon, \pi/2+\epsilon]$ 
yield the junction conditions of $F^{\mu\nu}$ in the form 
\begin{align}
	&[F^{\theta\nu}]^+_- =-4\pi j^\nu,
\label{eq:junction_j}
\\
	&[F_{r\nu}]^+_- = 0, 
\label{eq:junction_0}
\end{align}
where 
\begin{align}
	[F^{\mu\nu}]^+_- 
		:= \underset{\epsilon\to 0}{\rm lim}
	\left(
	F^{\mu\nu}|_{\theta=\pi/2+\epsilon}-F^{\mu\nu}|_{\theta=\pi/2-\epsilon}
	\right).
\end{align}

In order to separate the variables of the Maxwell equations in the metric \eqref{met:Kerr}, 
we use the Newman-Penrose formalism \cite{Teukolsky:1973ha}, in which 
$F_{\mu\nu}$ are represented by three complex scalars in the form 
\begin{align}
    \phi_0 \coloneqq &\ F_{\mu\nu}l^\mu m^\nu,\\
    \phi_1 \coloneqq &\ \frac{1}{2}F_{\mu\nu}(l^\mu n^\nu + m^{*\mu} m^\nu),\\
    \phi_2 \coloneqq &\ F_{\mu\nu}m^{*\mu} n^\nu.
  \end{align}
Here, $l^\mu,\ n^\mu$, $m^\mu$ and $m^{*\mu}$ are the null tetrad \cite{Kinnersley:1969zza}:
\begin{align}
    l^\mu =& \left(\frac{r^2+a^2}{\Delta},1,0,\frac{a}{\Delta}\right),\\
    n^\mu =& \frac{1}{2\Sigma}\left(r^2+a^2,-\Delta,0,a\right),\\
    m^\mu =& \frac{-\rho^*}{\sqrt{2}}\left(ia\sin\theta,0,1,\frac{i}{\sin\theta}\right),\\
    m^{*\mu} =& \frac{-\rho}{\sqrt{2}}\left(-ia\sin\theta,0,1,\frac{-i}{\sin\theta}\right),
\end{align}
where $\rho := -1/(r-ia\cos\theta)$ and ${}^*$ means complex conjugate. 
Conversely, the components $F^{\mu\nu}$ are represented by $\phi_0,\ \phi_1,\ \textrm{and}\ \phi_2$ as
\begin{align}
    F^{tr} = & 2\Re\left[\frac{r^2+a^2}{\Sigma}\phi_1
	+\frac{ia\rho^*\sin\theta}{\sqrt{2}}\left(\phi_2-\frac{\rho^2\Delta}{2}\phi_0\right)\right],
\label{Ftr}
\\
    F^{t\theta} =& 2\Re\left[\frac{ia\sin\theta}{\Sigma}\phi_1
	-\frac{r^2+a^2}{\sqrt{2}}\frac{\rho^{*}}{\Delta}
	\left(\phi_2-\frac{\rho^2\Delta}{2}\phi_0\right)\right],
\\
    F^{t\varphi}= & 2\Re\left[\frac{-i\rho^*\Sigma}{\sqrt{2}\Delta\sin\theta}
		\left(\phi_2 +\frac{\rho^2\Delta}{2}\phi_0\right)\right],
\label{E_phi}
\\
    F^{\theta\varphi} =& 2\Re\left[\frac{-i}{\Sigma\sin\theta}\phi_1
	+\frac{a\rho^*}{\sqrt{2}\Delta}
	\left(\phi_2-\frac{\rho^2\Delta}{2}\phi_0\right)\right],
\\
    F^{\varphi r} =&  2\Re\left[\frac{a}{\Sigma}\phi_1
	+\frac{i\rho^*}{\sqrt{2}\sin\theta}
	\left(\phi_2-\frac{\rho^2\Delta}{2}\phi_0\right)\right],
\\
    F^{r\theta} = & 2\Re\left[\frac{-\rho^*}{\sqrt{2}}
		\left(\phi_2+\frac{\rho^2\Delta}{2}\phi_0\right)\right].
\label{B_phi}
\end{align}

In the stationary and axisymmetric case, the Maxwell equations in vacuum are written as 
\begin{align}
    -\sqrt{2}\frac{\partial}{\partial r}\frac{\phi_1}{\rho^2} 
		=& \left(\frac{1}{\rho}\frac{\partial}{\partial\theta}+\frac{\cot\theta}{\rho}
	+ia\sin\theta\right)\phi_0,
\label{eq:reMax:l}
\\
    \sqrt{2}\frac{\partial}{\partial \theta}\frac{\phi_1}{\rho^2} 
		= &\left(\frac{\Delta}{\rho}\frac{\partial}{\partial r}+\frac{2(r-M)}{\rho}
		+\Delta\right)\phi_0,
\label{eq:reMax:m}
\\
    -\frac{\rho^3}{\sqrt{2}}\frac{\partial}{\partial \theta}\frac{\phi_1}{\rho^2} 
		= &\left(\frac{\partial}{\partial r}-\rho\right)\phi_2,
\label{eq:reMax:ms}\\
    \frac{\Delta\rho^2}{\sqrt{2}}\frac{\partial}{\partial r}\frac{\phi_1}{\rho^2} 
		= &\left(\frac{1}{\rho}\frac{\partial}{\partial\theta}
		-ia\sin\theta+\frac{\cot\theta}{\rho}\right)\phi_2.
\label{eq:reMax:n}
\end{align}

By elimination of $\phi_1$ from \eqref{eq:reMax:l} - \eqref{eq:reMax:n}, we obtain a couple of equations
\begin{align}
  	&\left(\frac{\partial}{\partial r}-\rho\right)
	\left(\phi_2+\frac{\Delta \rho^2}{2}\phi_0\right) =0,
\\
  	&\left(\frac{1}{\rho}\frac{\partial}{\partial\theta}
		+\frac{\cot\theta}{\rho}-ia\sin\theta\right)
		\left(\phi_2+\frac{\Delta \rho^2}{2}\phi_0\right) = 0. 
\end{align}
We can solve these in the form 
\begin{align}
	 \phi_2+\frac{\Delta \rho^2}{2}\phi_0 
		= C ~\frac{\rho}{\sin\theta},
\end{align}
where $C$ is an arbitrary constant. We should choose $C=0$, otherwise, the solution diverges at both $\theta= 0$ and $\theta=\pi$ on the rotation axis. Therefore, we have
\begin{align}
  \phi_2 = -\frac{\Delta \rho^2}{2}\phi_0, 
\label{sol:phi2}
\end{align}
and from  \eqref{E_phi} and \eqref{B_phi}, we obtain 
\begin{align}
	F^{t\varphi}=0, \quad \mbox{and}\quad F^{r\theta}=0.
\end{align}

On the other hand, from \eqref{eq:reMax:l} and \eqref{eq:reMax:m}, we derive  
the integrability condition for $\phi_1$ in the form \cite{Petterson:1975sg} 
\begin{align}
    \partial_r^2(\Delta \phi_0) 
	+\frac{1}{\sin\theta}\partial_\theta\left(\sin\theta\partial_\theta\phi_0\right)
	-\left(\frac{1}{\tan^2\theta}+1\right)\phi_0 = 0. 
\label{eq:Tue}
\end{align}
We solve \eqref{eq:Tue} for $\phi_0$, then $\phi_2$ is obtained from \eqref{sol:phi2}, 
and $\phi_1$ is determined from \eqref{eq:reMax:l} and \eqref{eq:reMax:m}.

\subsection{Vacuum solutions}\label{sec:sol_Max}

Setting $\phi_0 \coloneqq \Delta^{-1/2} R(r)\Theta(\theta)$, 
we can reduce \eqref{eq:Tue} into two ordinary differential equations
for $R(r)$ and $\Theta(\theta)$ in the form:
\begin{align}
  &\Delta\frac{d^2R(r)}{dr^2}+2(r-M)\frac{dR(r)}{dr}
	-\left(\lambda + \frac{M^2-a^2}{\Delta}\right) R(r) =0, 
\label{eq:asLgd}\\
  &\frac{d^2\Theta(\theta)}{d\theta^2}+\frac{1}{\tan\theta}\frac{d\Theta(\theta)}{d\theta}
	+\left(\lambda-\frac{1}{\sin^2\theta}\right) \Theta(\theta) = 0, 
\label{eq:asLgd2}
\end{align}
where $\lambda$ is a separation constant. 
Instead of $r$ and $\theta$, 
we introduce new variables: 
\begin{align}
	u \coloneqq \frac{r-M}{r_\mathrm{H}-M},
\quad \mbox{and}\quad
	x \coloneqq \cos\theta, 
\end{align}
where they vary in the range $1<u<\infty$ and $-1\leq x \leq 1$, 
then \eqref{eq:asLgd} and \eqref{eq:asLgd2} can be solved by 
the Legendre functions $P_\nu(u)$ and $P_\nu(x)$ in the form:
\begin{align}
  &R_\lambda(r) = \sqrt{u^2-1}\frac{dP_\nu(u)}{du} = \Delta^{1/2}\frac{dP_\nu(u)}{dr},
\label{sol:phi0_r}
\\
  &\Theta_\lambda (\theta) = \sqrt{1-x^2}\frac{dP_\nu(x)}{dx} = - \frac{dP_\nu(x)}{d\theta},
\label{sol:phi0_th}
\end{align}
where $\nu(\nu+1)=\lambda$. 

We require the radial function $R_\lambda(r)$ in \eqref{eq:asLgd} to be regular both at the horizon ($u=1$) and at infinity ($u\rightarrow\infty$). 
We assume that $\lambda=-k^2-1/4$, with continuous positive $k$, i.e., the Legendre functions with the complex indices $\nu =-1/2+ik$ (see \cite{Zhurina:1966gl}). 
Because $P_\nu(u)\simeq u^{-(4k^2+1)/8}$ as $u\rightarrow 1$, and $P_\nu(u)\simeq u^{-1/2+ ik}$ as $u\rightarrow\infty$, the solution described by \eqref{sol:phi0_r} is regular both at the horizon and at infinity. 

On the other hand, as for the angular function $\Theta_\nu(x)$, $P_\nu(x)$ is regular 
on the rotation axis in $\VN$, $x=1 (\theta=0)$, 
while singular on the axis in $\VS$, $x=-1 (\theta=\pi)$. 
Therefore, we adopt 
\begin{align}
  \phi_0 
	&= \frac{1}{\sqrt{2}}\int^\infty_0 dk\alpha_k \frac{dP_\nu(u)}{dr}
		\frac{dP_\nu (x)}{d\theta}, 
\label{sol:phi0}
\end{align}
as the solution to \eqref{eq:Tue} in $\VN$, where the coefficients $\alpha_k$ are complex. 
Integrating equations \eqref{eq:reMax:l} and \eqref{eq:reMax:m} for given $\phi_0$ by \eqref{sol:phi0}, we obtain 
\begin{align}
    \phi_1 =  \frac{\rho^2}{2} Q + \frac{\rho^2}{2}\int^\infty_0 dk \alpha_k S_k (u,x), 
\label{sol:phi1}
\end{align}
where
\begin{equation}
  S_k (u,x) := \nu(\nu+1)\frac{1}{\rho}P_\nu(u)P_\nu(x)
	-ia\sin\theta P_\nu(u)\frac{dP_\nu(x)}{d\theta}
	+\Delta\frac{dP_\nu(u)}{dr}P_\nu(x), 
\label{Sl}
\end{equation}
and $Q$ is a complex integral constant.

Next, we examine vacuum solutions in the region $\VS$. 
We replace $x$ in \eqref{sol:phi0} and \eqref{sol:phi1}
by $-x$, so that the vacuum solution in $\VS$ is given by
\begin{align}
	&\phi_0 
	= \frac{1}{\sqrt{2}}\int^\infty_0 dk\tilde\alpha_k \frac{dP_\nu(u)}{dr}
		\frac{dP_\nu (-x)}{d\theta}, 
\label{sol:south_phi_0}
\\
    &\phi_1 
	= \frac{\rho^2}{2} \tilde{Q}
	+ \frac{\rho^2}{2} \int^\infty_0 dk \tilde\alpha_k S_k (u,-x), 
\label{sol:south_phi_1}
\end{align}
where $\tilde{Q}$ is a complex constant, and $\tilde\alpha_k$ are complex coefficients. 
The southern vacuum solutions \eqref{sol:south_phi_0} and \eqref{sol:south_phi_1} are regular on the rotational axis in $\VS$ while singular on the axis in $\VN$. 

In this paper, 
we assume 
\begin{align}
	Q = \tilde{Q} = 0,
\label{eqs:Re:bc}
\end{align}
which means the central black hole has no electric nor magnetic charges (see appendix \ref{app:mn_srcs}). 

\section{magnetospheres with a thin disk model}\label{sec:bc}

We consider a magnetosphere with a disk current on a thin disk in the equatorial plane. 
In this section, from the junction conditions \eqref{eq:junction_j} and \eqref{eq:junction_0} 
we determine the coefficients $\alpha_k$ and $\tilde{\alpha}_k$.
Hereafter, we use the abbreviation 
\begin{align}
	X|_\pm : = \lim_{\epsilon\to 0} X(\theta)\big|_{\theta=\frac{\pi}{2}\pm \epsilon}
\end{align}
for any $X$. Noting that 
\begin{align}
	&P_\nu(x)|_-=P_\nu(-x)|_+, \quad 
	\frac{dP_\nu(x)}{d\theta}\Big|_- =-\frac{dP_\nu(-x)}{d\theta}\Big|_+ , 
\end{align}
\begin{align}
  &S^\Re_{k-}= S^\Re_{k+} = \left( -\nu(\nu+1) r P_\nu(u)	
			+\Delta\frac{dP_\nu(u)}{dr}\right)P_\nu(0), 
\cr
  &S^\Im_{k-} 
	= -aP_\nu(u)\frac{dP_\nu(x)}{d\theta}\Big|_-
  =-S^\Im_{k+} 
	= aP_\nu(u)\frac{dP_\nu(-x)}{d\theta}\Big|_+, 
\end{align}
we have
\begin{align}
	[ \phi^\Re_{1}]^+_- 
	&= \frac{1}{2r^2} \int^\infty_0 dk \left(\tilde{\alpha}_k^\Re S^\Re_{k+} -\tilde{\alpha}_k^\Im S^\Im_{k+} 
		- \alpha_k^\Re S^\Re_{k-} +\alpha_k^\Im S^\Im_{k-} \right)
\cr
	&= \frac{1}{2r^2}  \int^\infty_0 dk(-\alpha_k^\Re + \tilde{\alpha}_k^\Re) S^\Re_{k-} 
	+\frac{1}{2r^2}  \int^\infty_0 dk(\alpha_k^\Im + \tilde{\alpha}_k^\Im)S^\Im_{k-},
\cr
	[\phi^\Im_{1}]^+_- 
	&= \frac{1}{2r^2}  \int^\infty_0 dk \left(\tilde{\alpha}_k^\Im S^\Re_{k+} +\tilde{\alpha}_k^\Re S^\Im_{k+} 
		- \alpha_k^\Im S^\Re_{k-} -\alpha_k^\Re S^\Im_{k-} \right)
\cr
	&= \frac{1}{2r^2}  \int^\infty_0 dk (-\alpha_k^\Im+ \tilde\alpha_k^\Im) S^\Re_{k-} 
	+ \frac{1}{2r^2}  \int^\infty_0 dk (-\alpha_k^\Re - \tilde\alpha_k^\Re)S^\Im_{k-},
\end{align}
and
\begin{align}
	[\phi^\Re_{0}]^+_-
	&=\frac{1}{\sqrt{2}}\int^\infty_0 dk\left( \tilde{\alpha}_k^\Re \frac{dP_\nu(u)}{dr}\frac{dP_\nu}{d\theta}\Big|_{+} 
	- \alpha_k^\Re \frac{dP_\nu(u)}{dr}\frac{dP_\nu}{d\theta}\Big|_{-} \right)
\cr
	&=\frac{1}{\sqrt{2}}\int^\infty_0 dk(-\alpha_k^\Re - \tilde\alpha_k^\Re) 
	\frac{dP_\nu(u)}{dr}\frac{dP_\nu}{d\theta}\Big|_{-},
\cr	
	[\phi^\Im_{0}]^+_-
	&= \frac{1}{\sqrt{2}}\int^\infty_0 dk \left(\tilde{\alpha}_k^\Im \frac{dP_\nu(u)}{dr}\frac{dP_\nu}{d\theta}\Big|_{+} 
	- \alpha_k^\Im \frac{dP_\nu(u)}{dr}\frac{dP_\nu}{d\theta}\Big|_{-} \right)
\cr
	&= \frac{1}{\sqrt{2}}\int^\infty_0 dk(-\alpha_k^\Im - \tilde\alpha_k^\Im )
		\frac{dP_\nu(u)}{dr}\frac{dP_\nu}{d\theta}\Big|_{-},
\end{align}
where the superscript $\Re,\Im$ stand for the real and imaginary parts, respectively.

 \subsection{Junction conditions}
First, the junction condition \eqref{eq:junction_0} leads to
\begin{align}
	&[F_{r\varphi}]^+_- =-2 a[\phi_1^\Re]^+_- 
		+ 2\frac{r^2+a^2}{\sqrt{2}r}[\phi_0^\Im]^+_-
	=0,\label{jnc:no_mg_ch1}
\\
	&[F_{tr}]^+_- =-2 [\phi_1^\Re]^+_-
		+2 \frac{a}{\sqrt{2}r}[\phi_0^\Im]^+_-
	=0.\label{jnc:no_mg_ch2}
\end{align}
Therefore we obtain
\begin{align}
	[\phi_1^\Re]^+_-=0,\quad\mbox{and}\quad [\phi_0^\Im]^+_-=0,
\end{align}
thus, 
\begin{align}
	\alpha^\Re_k =\tilde\alpha^\Re_k ,\quad\mbox{and}\quad  
	\alpha^\Im_k =-\tilde\alpha^\Im_k.
\label{alpha_symm}
\end{align}
In addition to \eqref{jnc:no_mg_ch1} and \eqref{jnc:no_mg_ch2}, 
the conditions \eqref{alpha_symm} require that $F^{\mu\nu}$ satisfy 
\begin{equation}
	F_{\theta t}|_- = -F_{\theta t}|_+,\quad F_{\theta\varphi}|_- = - F_{\theta\varphi}|_+
\end{equation}
on the equatorial plane.

Next, the junction condition \eqref{eq:junction_j} leads to 
\begin{align}
	&[F^{\theta\varphi}]^+_-
	 =\frac{2}{r^2}[ \phi_1^\Im]^+_- + 2\frac{a}{\sqrt{2}r^3}[\phi_0^\Re]^+_-
	=-4\pi j^\varphi,
\label{junction_current}
\\
	&[F^{t\theta}]^+_- 
	=-\frac{2a}{r^2}\left[ \phi_1^\Im\right]^+_- -2\frac{r^2+a^2}{\sqrt{2}r^3} [\phi_0^\Re]^+_-
	=4\pi j^t,
\label{junction_charge}
\end{align}
where $j^\varphi$ and $j^t$ are toroidal current and charge surface density on the equatorial plane. In this paper, we assume $[\phi_0^\Re]^+_-=0$, i.e., 
\begin{align}
	\alpha^\Re_k =\tilde\alpha^\Re_k=0,
\label{alpha_re_zero}
\end{align}
for simplicity. We see that $j^\varphi$ and $j^t$ are related as
\begin{align}
	j^t=aj^\varphi.
\label{jphi-jt}
\end{align}
In our model, the charge surface density is induced on the disk by the rotation of the black hole. We note that $j^\mu$ is a spacelike vector regardless of the spin $a$. 

\subsection{Electric current model on the equatorial plane}

We will specify the current and charge distribution on the thin disk with the inner edge as a model, and obtain electromagnetic fields produced by the source.
It is difficult to determine $\alpha^\Im_k$ from the junction condition \eqref{junction_current} with \eqref{alpha_re_zero} directly because $\phi_1^\Im$ include both $P_\nu(u)$ and its derivative. 
Multiplying the junction condition \eqref{junction_current} by $r^4$
and differentiating with respect to $r$, we obtain  
\begin{align}
	4\pi \frac{d}{dr} (r^4 j^\varphi)  
	=&-2\frac{d}{dr}\left[ r^2 \phi_1^{\Im}\right]^+_- \notag\\
    =&2 r\frac{d}{dr}\int_0^\infty dk (-\nu(\nu+1))\alpha^\Im_k P_\nu(0)P_\nu(u),
\label{eq:equality}
\end{align}
where, in the second equality, we have used the expansion of $\phi_1^{\Im}$ given by \eqref{sol:phi1} 
and \eqref{Sl} with the parameters choice in \eqref{eqs:Re:bc}, \eqref{alpha_symm} 
and \eqref{alpha_re_zero}. 
Here, rewrite the left hand side of \eqref{eq:equality} as 
\begin{align}
	2\pi \frac{d}{dr} (r^4 j^\varphi) 
	= r\frac{d}{dr}\mathcal{I}(u),
\label{Ijrelation}
\end{align}
using a function $\mathcal{I}(u)$. Integrate \eqref{eq:equality} with \eqref{Ijrelation}, we obtain
\begin{align}
  \mathcal{I}(u)
	=& \int_0^\infty dk (-\nu(\nu+1))\alpha^\Im_k P_\nu(0)P_\nu(u),
\label{def_I}
\end{align}
where we set the integration constant to be zero. 
The function ${\cal I}(u)$ gives the toroidal current distribution $j^\varphi$ on the equatorial plane. 
Simultaneously, it provides the electric charge distribution $j^t$ from \eqref{jphi-jt}. 

Fortunately, there exists the inversion of the integral \eqref{def_I} 
\begin{align}
    \alpha^\Im_k 
	= -\frac{k\tanh (k\pi)}{\nu(\nu+1)P_\nu(0)}\int_1^\infty du \mathcal{I}(u)P_\nu(u), 
\label{eqs:inv_fml}
\end{align}
which is known as the Mehler-Fock transformation \cite{Bateman:1953ge,Zhurina:1966gl}.
When ${\cal I}(u)$ is given, we obtain the explicit form of $\alpha^\Im_k$, therefore, $\phi_0,\ \phi_1,\textrm{and}\ \phi_2$ are determined.

Here, we set ${\cal I}(u)$ explicitly and give a model of current distribution 
on the thin disk with an inner edge at $u = u_0$. 
Let us assume 
\begin{align}
    \mathcal{I}(u) = \left\{
		\begin{array}{ll}
        0 & \quad \mbox{for}\quad 1\leq u<u_0 \\ 
        \mathcal{I}_0Q_s(u)&\quad \mbox{for}\quad u_0 \leq u
    \end{array}\right. ,
\label{anz:src}
\end{align}
where $\mathcal{I}_0$ is an arbitrary real constant and $Q_s(u)$ is the Legendre function 
of second kind. 
We obtain the explicit form of $\alpha^\Im_k$ through the integral formula 
involving two Legendre functions.  
Since the Legendre function $Q_s(u)$ falls off in the form $u^{-s-1}$, 
the index $s$ should be chosen as a positive number so that ${\cal I}(u)$ decreases to zero 
at large $u$.

Integrating \eqref{Ijrelation} with \eqref{anz:src}, we obtain
\begin{align}
	2\pi r^4j^\varphi =\frac{2\pi }{a}r^4 j^t
	= \left\{
		\begin{array}{cl}
    	0 & \quad \mbox{for}\quad 1\leq u<u_0 \\
    	\displaystyle \mathcal{I}_0 r Q_s(u)
		-\frac{\mathcal{I}_0}{s(s+1)}\Delta \frac{dQ_s(u)}{dr}
		+\left.\frac{\mathcal{I}_0}{s(s+1)}\Delta\frac{dQ_s(u)}{dr}\right|_{u=u_0}
		&\quad \mbox{for}\quad u_0 \leq u 
  \end{array}\right. .
\label{sorc:phi1}
\end{align}
We can check $4\pi j^\nu = - [F^{\theta\nu}]^+_- =0$ in the range $1\leq u<u_0$ (see appendix \ref{app:gap}), and we must choose an integration constant such that the total current $\int^\infty_{r_\mathrm{H}} dr r^2 j^\varphi = 0$ (see appendix \ref{app:mn_srcs}).

We obtain the coefficients $\alpha^\Im_k$ by \eqref{eqs:inv_fml} for the source \eqref{sorc:phi1} in the form
\begin{align}
  \alpha^\Im_k 
	= -\frac{1}{P_\nu(0) (k^2+\frac{1}{4})}~
	\frac{\mathcal{I}_0k\tanh(\pi k)}{k^2+(s+\frac{1}{2})^2}(u_0^2-1)
	\left(P_\nu(u)\frac{dQ_s(u)}{du}-\frac{dP_\nu(u)}{du}Q_s(u)\right)\Big|_{u=u_0}.
\label{def:alp}
\end{align}

From \eqref{Ftr} - \eqref{B_phi} with \eqref{alpha_symm} and \eqref{alpha_re_zero}, the nonvanishing components of $F^{\mu\nu}$ in the disk current model \eqref{sorc:phi1} are 
\begin{align}
    F^{tr} 
	=  & \frac{1}{\Sigma}\int dk\alpha^\Im_k\Biggl\{\frac{r^2+a^2}{\Sigma}
	\left[\nu(\nu + 1) P_\nu (u)-\frac{2r\Delta}{\Sigma}\frac{dP_\nu (u)}{dr}\right]a\cos\theta P_\nu (\cos\theta) 
\cr
    &\hspace{28truemm} +\left[\frac{r^2+a^2}{\Sigma}\frac{r^2-a^2\cos^2\theta}{\Sigma}P_\nu (u) 
	-\frac{r\Delta}{\Sigma}\frac{dP_\nu (u)}{dr}\right] a\sin\theta\frac{dP_\nu (\cos\theta)}{d\theta}\Biggr\}
\label{sol:Ftr}
\\
     F^{t\theta} 
	= &\frac{1}{\Sigma} \int dk \alpha^\Im_k\Biggl\{\frac{a\sin\theta}{\Sigma}
	\left[\nu(\nu + 1) rP_\nu (u)-\frac{r^2-a^2\cos^2\theta}{\Sigma}\Delta\frac{dP_\nu (u)}{dr}\right]P_\nu (\cos\theta)
\cr
    &\hspace{28truemm} 
	+  \left[-\frac{2a^2r\sin^2\theta}{\Sigma^2}P_\nu (u)
	+\frac{r^2+a^2}{\Sigma}\frac{dP_\nu (u)}{dr}\right]a\cos\theta\frac{dP_\nu (\cos\theta)}{d\theta} \Biggr\}
\label{sol:Ftth}
\\
    F^{r\varphi} 
		= & \frac{1}{\Sigma}\int dk \alpha^\Im_k\Biggl\{\frac{a^2\cos\theta}{\Sigma}
	\left[-\nu(\nu + 1) P_\nu (u) + 2r\frac{\Delta}{\Sigma}\frac{dP_\nu (u)}{dr}\right]
	P_\nu (\cos\theta)
\cr
    &\hspace{28truemm} 
	+ \left[-\frac{a^2\sin\theta}{\Sigma}\frac{r^2-a^2\cos^2\theta}{\Sigma}P_\nu (u)  
	+ \frac{r}{\sin\theta}\frac{\Delta}{\Sigma}\frac{dP_\nu (u)}{dr}\right]
	\frac{dP_\nu (\cos\theta)}{d\theta} \Biggr\}
\label{sol:Frph}
\\
    F^{\theta\varphi} 
	=& \frac{1}{\Sigma}\int dk \alpha^\Im_k\Biggl\{\frac{1}{\Sigma\sin\theta}\left[-\nu(\nu + 1) rP_\nu(u) 
	+ \frac{r^2-a^2\cos^2\theta}{\Sigma}\Delta\frac{dP_\nu (u)}{dr}\right]
	P_\nu (\cos\theta)
\cr
    &\hspace{28truemm} 
	+  \left[\frac{2r}{\Sigma}P_\nu (u) 
	- \frac{dP_\nu (u)}{dr}\right]\frac{a^2\cos\theta}{\Sigma}\frac{dP_\nu (\cos\theta)}{d\theta} \Biggr\},
\label{sol:Fthph}
\end{align}
where the coefficients $\alpha^\Im_k$ are given by \eqref{def:alp}. 

The magnetic flux function $\Psi$ for stationary and axisymmetric electromagnetic fields is defined by
\begin{align}
	\Psi(r,\theta) := \int_{0}^{\theta} F_{\theta\varphi }(r,\tilde{\theta})d\tilde{\theta} \label{mg_flx1},
\end{align}
where the integration is done along the curve $r=const., \varphi=const.$ The electric potential $\Phi$ is defined by
\begin{align}
	\Phi(r,\theta) := -\int_\infty^{r} F_{rt}(\tilde{r},\theta) d\tilde{r}, \quad\label{scl_ptn1}
\end{align}
where the integration is done along the line $\theta=const.,\varphi=const.$ It is obvious that $\Phi(\infty)=0$. From \eqref{sol:Ftr} - \eqref{sol:Fthph}, we obtain $\Psi$ and $\Phi$ explicitly as 
\begin{align}
    \Psi(r,\theta) = & 
	 \int_0^\infty dk\alpha^\Im_k
	\Biggl[-a^2\cos\theta\sin^2\theta\frac{\Delta}{\Sigma}\frac{dP_\nu}{dr}P_\nu(\cos\theta)
\notag\\
    &\hspace{40mm} 
	+ r\sin\theta\frac{r^2+a^2}{\Sigma}P_\nu(u)\frac{dP_\nu(\cos\theta)}{d\theta}
	-\Delta\frac{\sin\theta}{\nu(\nu + 1)}\frac{dP_\nu}{dr}\frac{dP_\nu(\cos\theta)}{d\theta}\Biggr],
\label{mg_flx2}\\
    \Phi(r,\theta) = & 
	\int_0^\infty dk\alpha^\Im_k
	\left[-a\cos\theta \frac{\Delta}{\Sigma}\frac{dP_\nu}{dr}P_\nu(\cos\theta)
	+\frac{ra\sin\theta}{\Sigma}P_\nu(u)\frac{dP_\nu(\cos\theta)}{d\theta}\right]. 
\label{scl_ptn2}
\end{align}

On the horizon, we see that 
\begin{equation}
	\Phi(r_\mathrm{H},\theta) = \omega_\mathrm{H}\Psi(r_\mathrm{H},\theta) =\omega_\mathrm{H} \sin\theta\frac{2Mr_\mathrm{H}^2}{\Sigma_\mathrm{H}}\int_0^\infty dk\alpha^\Im_k \frac{dP_\nu(\cos\theta)}{d\theta}
	,\label{eq:PhivsPsi}
\end{equation}
where $\omega_\mathrm{H}$ is the angular velocity of the black hole at the horizon, defined by $\omega_\mathrm{H} = -g_{t\varphi}/g_{\varphi\varphi}|_{r=r_\mathrm{H}} = a/(2Mr_\mathrm{H})$, and $\Sigma_\mathrm{H} = r_\mathrm{H}^2+a^2\cos^2\theta$. 
The $\theta$ dependence of $\Phi$ and $\Psi$ on the horizon is identical, i.e., the ratio $\Phi$ to $\Psi$ is constant on the horizon. 
While the $\theta$ dependence is determined by $\alpha^\Im_k$ that are given by the current model, the ratio $\omega_\mathrm{H}$ is independent of the model. 
The first equality in \eqref{eq:PhivsPsi} is derived from the regularity of electromagnetic fields on the horizon \cite{Carter:1971zc,Lasota:2013kia}.

For convenience in considering motions of charged particles, we note that the gauge potential can be chosen as
\begin{align}
	A_t = -\Phi, \quad A_\varphi = \Psi, \quad A_r=A_\theta=0. 
\end{align} 

\newpage

\section{properties of electromagnetic fields in a thin disk model}
\label{sec:num_any}

\begin{figure}[H]
  \centering \includegraphics[width=0.7\columnwidth]{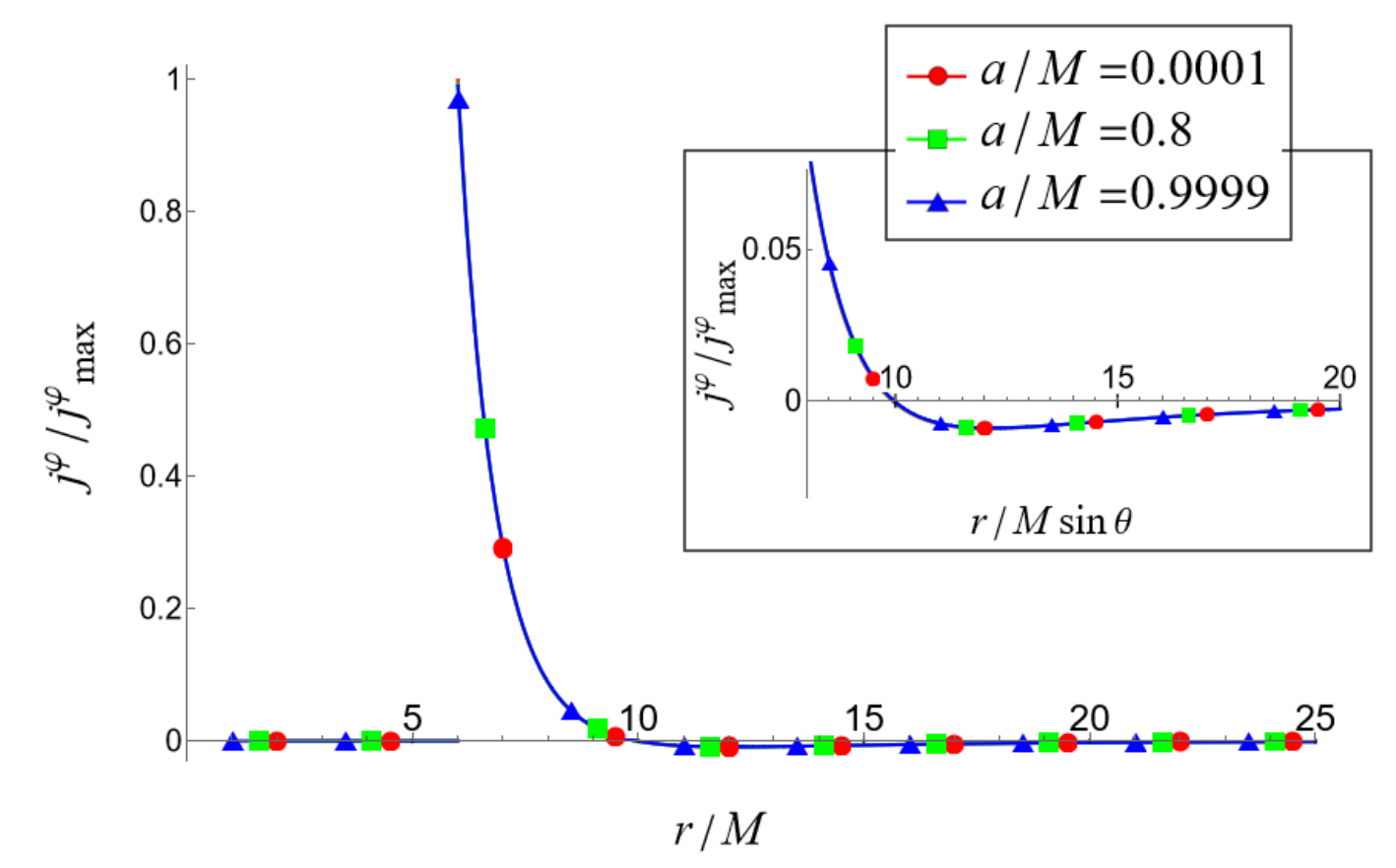}
\caption{%
The surface toroidal current density on the disk, $j^\varphi$, given by \eqref{sorc:phi1} 
with $s=2$ and $r_0/M=6$ is shown as functions of $r/M$ for the cases $a/M=0.0001,\ 0.8, \mbox{and }0.9999$. The current distributions for all spins are almost overlapping. 
}\label{fig:crnt_diff_a}
\end{figure}

Now we show the solutions graphically to study properties of electromagnetic fields such as the contour of $\Phi,\ \Psi$ and the distribution of $F^{\mu\nu}$. By using the definitions of the three-dimensional quantities, electric fields $\E$, electric flux densities $\D$, and magnetic flux densities $\B$, given in Appendix \ref{app:EM_in_ZAMO} in addition to $\Psi$ and $\Phi$, we see how the fields depend on the spin parameter $a$.
We fix the inner edge of the thin disk as $r_0/M = 6$ for all cases of $a$. 
In this choice, because $u_0 =(r_0-M)/\sqrt{M^2-a^2}\geq 5$, the current given by \eqref{sorc:phi1} behaves as $j^\varphi \sim r^{-4}$ independently of $a$ in the range $u_0<u$. 
The current distributions given by \eqref{sorc:phi1} with $s=2$ for different values of $a$ are shown in \figref{fig:crnt_diff_a}. 
The surface charge density, $j^t$, is given by \eqref{jphi-jt}. 
In our models, the direction of the toroidal current reverses at a radius $r/M\sim 10 $ on the disk.

\subsection{Magnetic field}\label{sec:mag_field}

\begin{figure}[H]
	\centering\subfloat[$a/M=0.0001$\quad(global veiw) \label{subfig:mag_flx_a00_out}]{%
    \includegraphics[width=0.35\columnwidth]{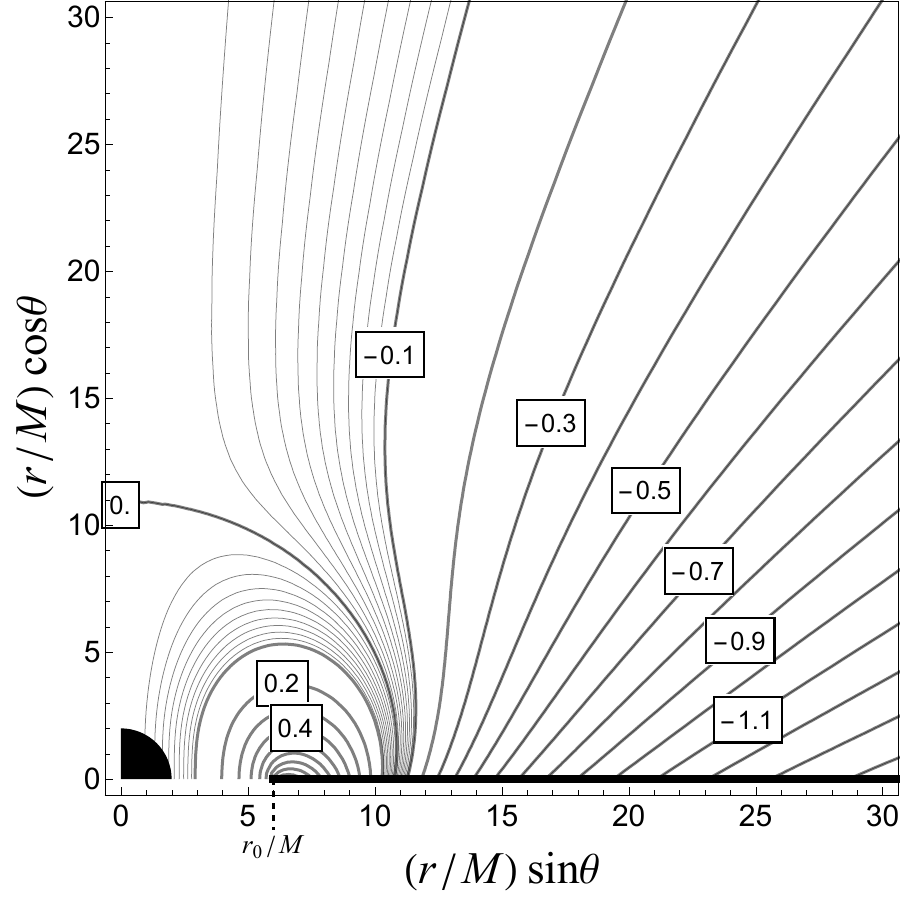}}\quad
    \subfloat[$a/M=0.0001$\quad(near horizon) \label{subfig:mag_flx_a00}]{%
    \includegraphics[width=0.35\columnwidth]{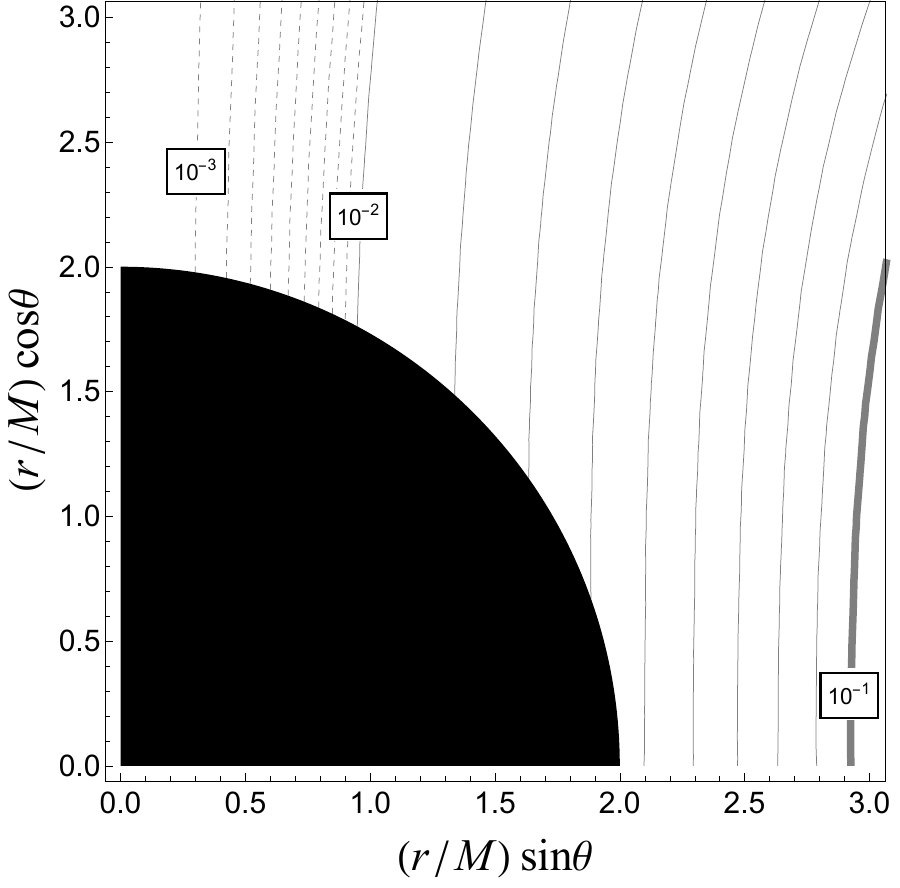}}

  \subfloat[$a/M=0.8$\quad(global veiw)\label{subfig:mag_flx_a08_out}]{%
    \includegraphics[width=0.35\columnwidth]{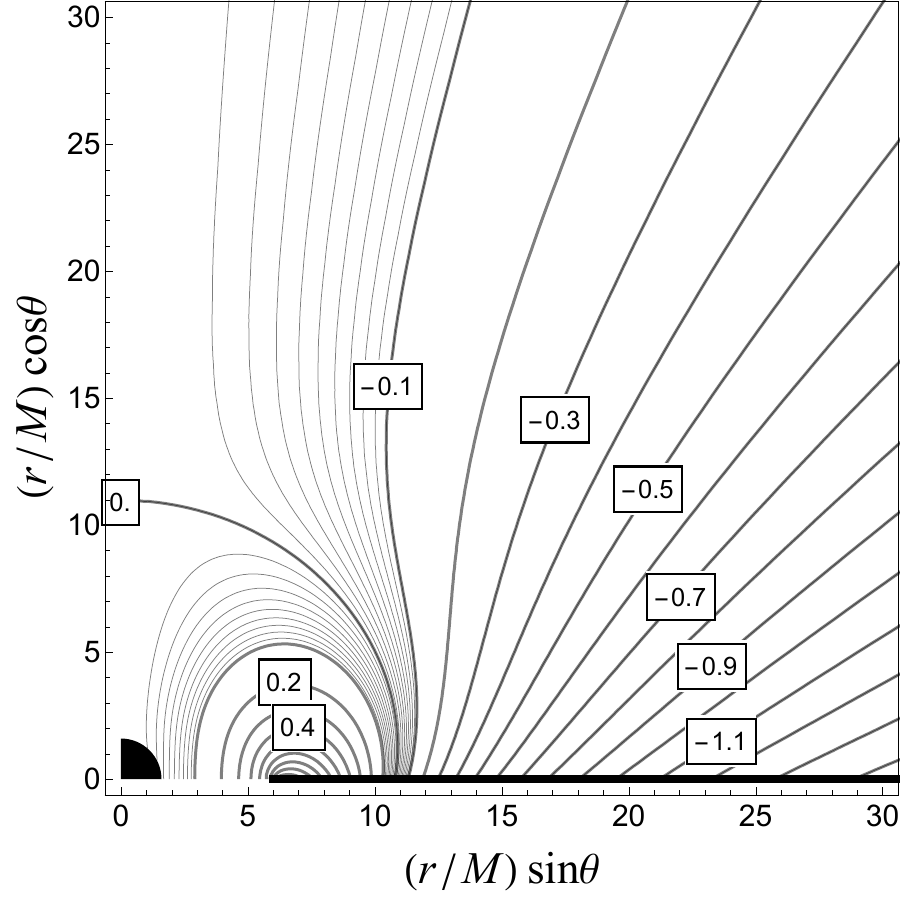}%
    }\quad\subfloat[$a/M=0.8$\quad(near horizon)\label{subfig:mag_flx_a08}]{%
    \includegraphics[width=0.35\columnwidth]{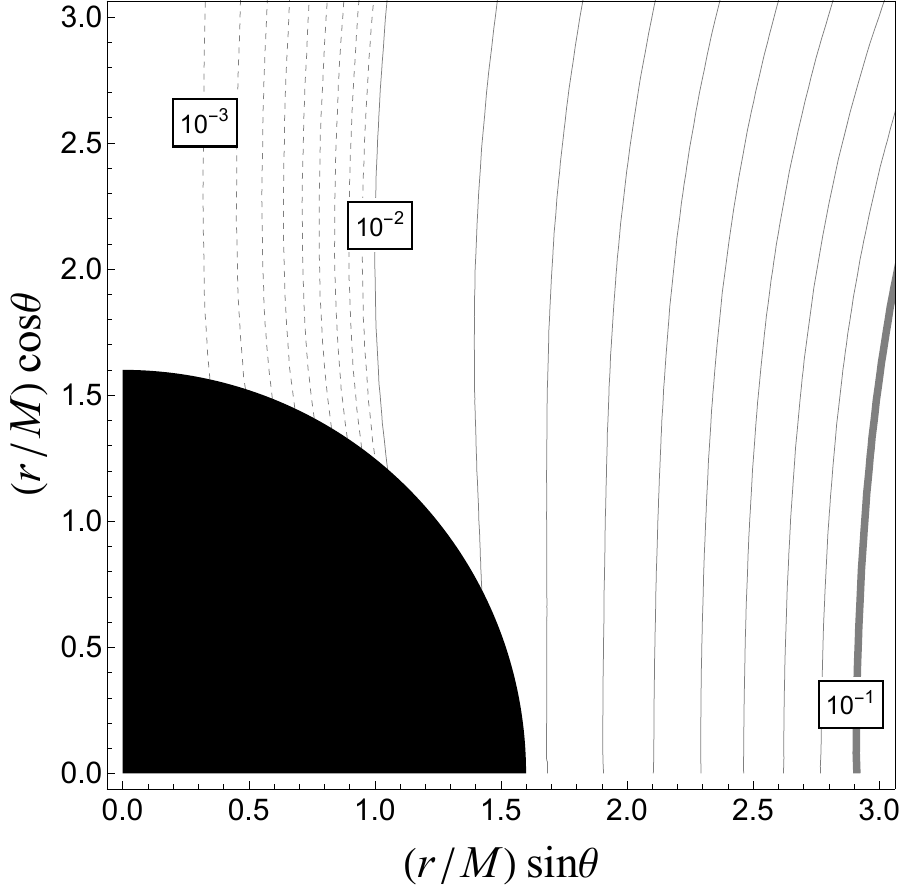}%
    }

  \subfloat[$a/M=0.9999$\quad(global veiw)\label{subfig:mag_flx_a09d4_out}]{%
    \includegraphics[width=0.35\columnwidth]{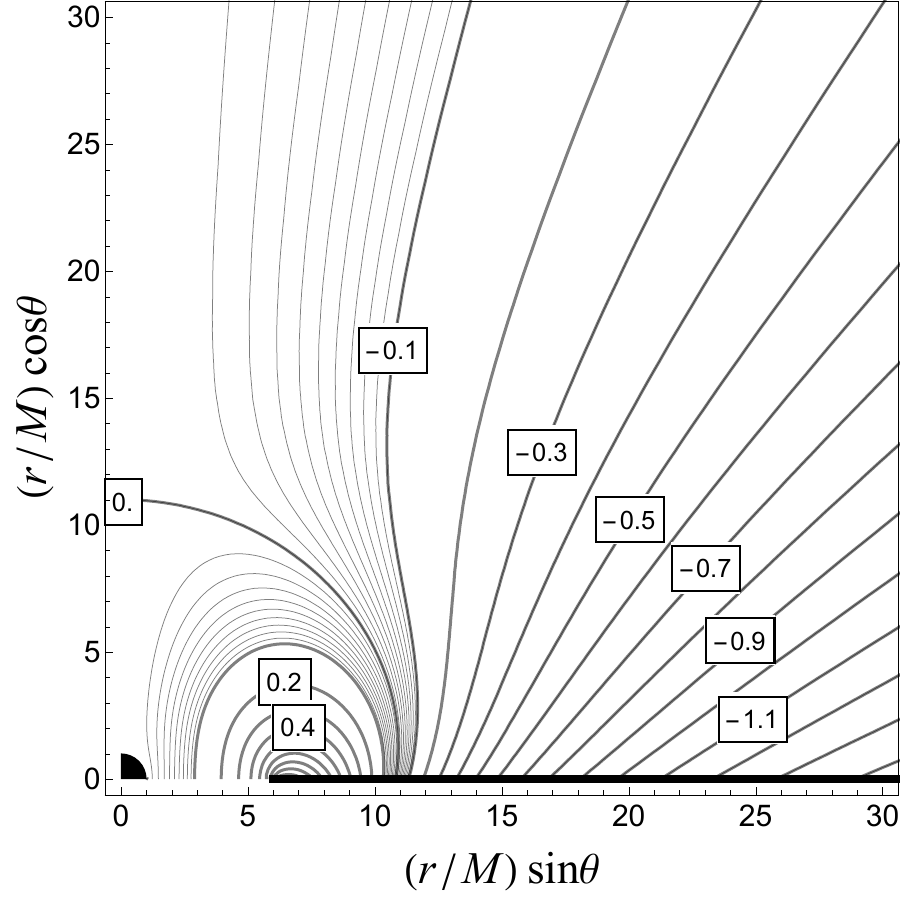}%
    }\quad
    \subfloat[$a/M=0.9999$\quad(near horizon)\label{subfig:mag_flx_a09d4}]{%
    \includegraphics[width=0.35\columnwidth]{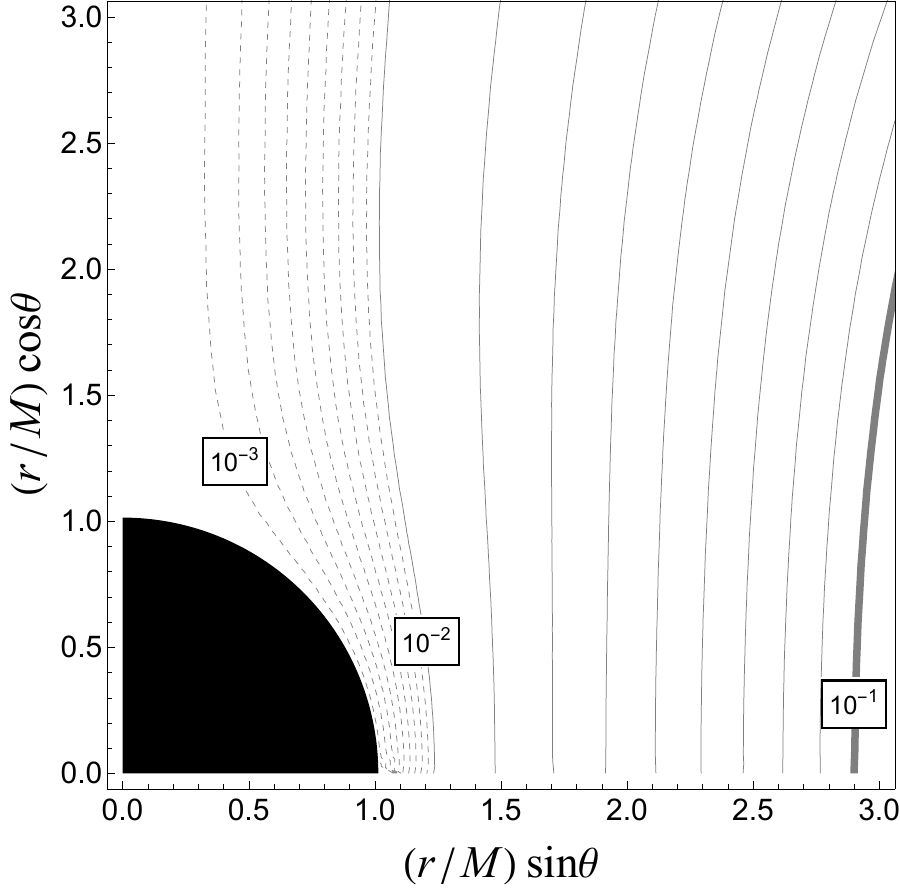}%
    }
  
  \caption{
Contours of the magnetic flux function $\Psi/\Psi^\mathrm{max}$, that is, magnetic flux lines, 
for the cases \protect\subref{subfig:mag_flx_a00_out},\protect\subref{subfig:mag_flx_a00} $a/M=0.0001$; \protect\subref{subfig:mag_flx_a08_out},\protect\subref{subfig:mag_flx_a08} $a/M=0.80$; \protect\subref{subfig:mag_flx_a09d4_out},\protect\subref{subfig:mag_flx_a09d4} $a/M=0.9999$, are shown. The quantity $\Psi^\mathrm{max}$ is almost the same, regardless of the spin $a$. 
The global views are shown in the left panels \protect\subref{subfig:mag_flx_a00_out},\protect\subref{subfig:mag_flx_a08_out},\protect\subref{subfig:mag_flx_a09d4_out}, and the close-up views near the horizon are in the right panels \protect\subref{subfig:mag_flx_a00},\protect\subref{subfig:mag_flx_a08},\protect\subref{subfig:mag_flx_a09d4}. 
In each panel, the black region shows the black hole, and the thick line on the horizontal axis shows the thin disk on the equatorial plane. 
}\label{fig:mag_flx}
\end{figure}

First, in \figref{fig:mag_flx}, we show the magnetic flux lines, integral curves of the magnetic flux density $\B$, that are given by the contours of the magnetic flux function $\Psi(r,\theta)$ defined in \eqref{mg_flx2}. 
We find, in \figsref{fig:mag_flx} \subref{subfig:mag_flx_a00_out}, \subref{subfig:mag_flx_a08_out}, and \subref{subfig:mag_flx_a09d4_out}, the $(r, \theta)$ plane is divided into two regions by the contour $\Psi=0$ that 
connects the disk and the zero point of the magnetic flux density $\B$ on the rotation axis. We refer to the $\Psi=0$ curve as the \lq separatrix\rq\ curve. 
The position of the zero point, $r/M=r_\mathrm{sep}/M\sim 11$ on the axis in the present model, 
does not depend on the spin parameter $a$. 
Because of the axisymmetry, the separatrix forms an almost spherical surface with the radius $r\sim r_\mathrm{sep}$ that divides the magnetosphere into two parts: the inner magnetosphere, and the outer one. 
In the inner region, we see two types of magnetic flux lines: loop-shaped lines surrounding the inner edge of the disk, and lines connecting the disk and the horizon. 
On the other hand, in the outer region, field lines connect the disk and the distant region almost radially. 
These behaviors are the same in all cases of different values of $a$. 
The separatrix exists because the direction of the current on the disk is reversed, as shown in \figref{fig:crnt_diff_a}. $\Psi^\mathrm{max}$ is the value of $\Psi$ at the point located near the inner edge on the disk, and $\Psi^\mathrm{max}$ is almost the same value regardless of the spin.

Close-up views of the magnetic flux lines near the horizon is shown in \figsref{fig:mag_flx} 
\subref{subfig:mag_flx_a00}, \subref{subfig:mag_flx_a08}, and \subref{subfig:mag_flx_a09d4}.  
In the cases \subref{subfig:mag_flx_a00} $a/M=0.0001$ and \subref{subfig:mag_flx_a08} $a/M=0.8$, the field lines are almost parallel to the rotation axis and they penetrate the horizon, while in the case \subref{subfig:mag_flx_a09d4} $a/M=0.9999$, where the spin parameter approaches its maximum value, the magnetic flux lines are expelled from the event horizon, that is, $\Psi\to 0$ on the horizon. 
This behavior is called the black hole Meissner effect \cite{Chamblin:1998qm,Bicak:1976ag,Bicak:1985af}. 

In short, the global behavior of the magnetic field is almost the same as in the case of TT01, and the behavior of the field near the horizon is the 
same as that described by Wald's solutions \cite{Wald:1974np}.

\subsection{Electric field}
\label{sec:ele_pot}
Next, we show the electric field in our model. 
In a Schwarzschild spacetime, solutions of the magnetic field without an electric field are allowed (TT01). 
On the other hand, in a Kerr spacetime, the electric potential $\Phi$ in \eqref{scl_ptn2} is non-vanishing if $a\neq 0$, namely, electric field appears due to the rotation of the black hole. 
This fact holds for any choice of the coefficients $\alpha^\Im_k$, which are determined by the current on the disk. 
From \eqref{scl_ptn2}, it is obvious that the potential value at the north pole on the horizon is equal to the potential at infinity, namely,
\begin{align}
	\Phi(r_\mathrm{H}, 0)=\Phi(\infty)=0. 
\end{align}
This is also true for any choice of $\alpha^\Im_k$. 
In Wald's solutions, the electric field also appears without a current source, and the potential difference between the north pole on the horizon and infinity arises (see appendix \ref{app:wald}).

In \figref{fig:scl_pot}, we plot the contours of $\Phi$ for $a/M = 0.0001,\ 0.8,\ 0.9999$, respectively. 
We see that a saddle point of $\Phi$ appears on the rotation axis at $r/M=r_\mathrm{sad}/M \sim 5$ no matter how $a$ takes different values. 
We should note that the position of the saddle point of $\Phi$ does not coincide with the zero point of $\B$.  
In the region $r \gtrsim r_\mathrm{sad}$, the shape of the contours of $\Phi$ is almost the same in spite of different values of $a$, while the norm of the gradient of $\Phi$, i.e., $|\E|$, is proportional to $a$ even if $a/M\sim 1$ as shown in \figref{fig:dep_a:At}. However, near the horizon, the shape varies with the spin, and $|\E|$ is not proportional to the spin. 

We see, in \figsref{fig:scl_pot} \subref{subfig:sca_pot_a00}, \subref{subfig:sca_pot_a08}, and \subref{subfig:sca_pot_a09d4}, the shape of $\Phi$ depends on $a$ in the vicinity of the horizon. 
In the cases of $a/M = 0.0001$ and $0.8$, contours of $\Phi$ intersect with the horizon, 
while in the case of $a/M =0.9999$, near the maximum value, 
the horizon becomes the equipotential surface $\Phi=0$. 
From this result and the black hole Meissner effect, we see that in the limit $a/M \to 1$, $\Psi\to 0$ and $\Phi\to 0$ occur on the horizon, simultaneously, as shown analytically in \eqref{eq:PhivsPsi}.

\begin{figure}[H]
  \centering\subfloat[$a/M=0.0001$\quad(global veiw)\label{subfig:sca_pot_a00dis}]{%
  \includegraphics[width=0.35\columnwidth]{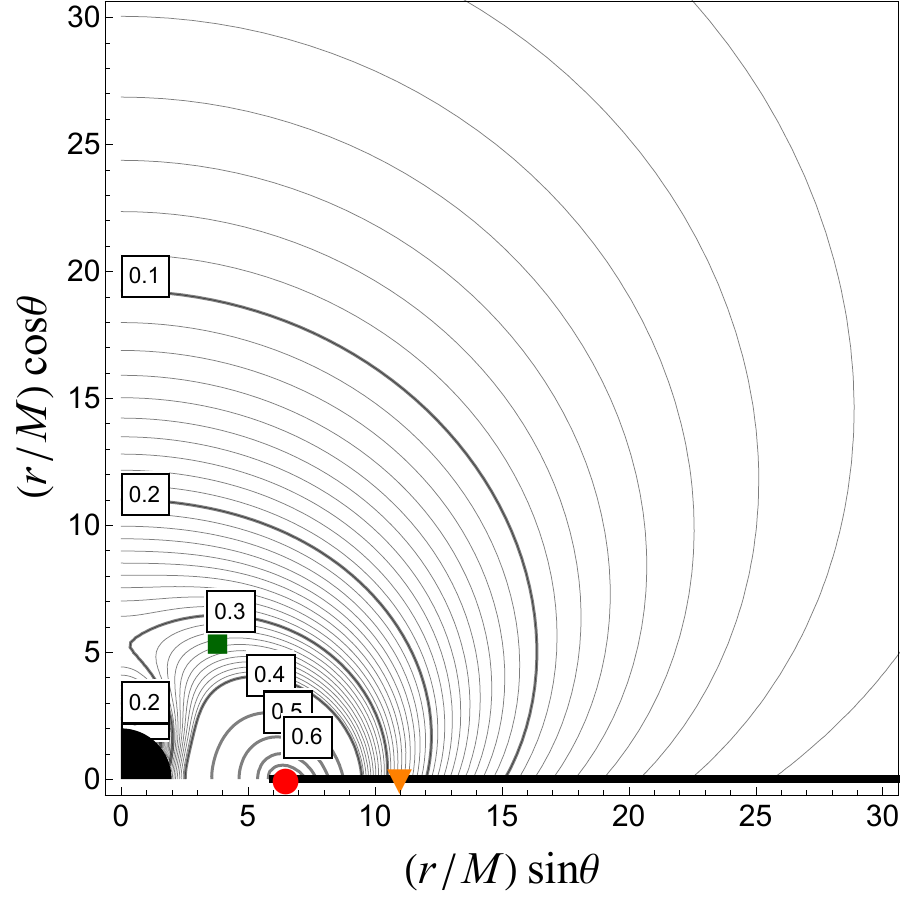}%
  }\quad\subfloat[$a/M=0.0001$\quad(near horizon)\label{subfig:sca_pot_a00}]{%
  \includegraphics[width=0.35\columnwidth]{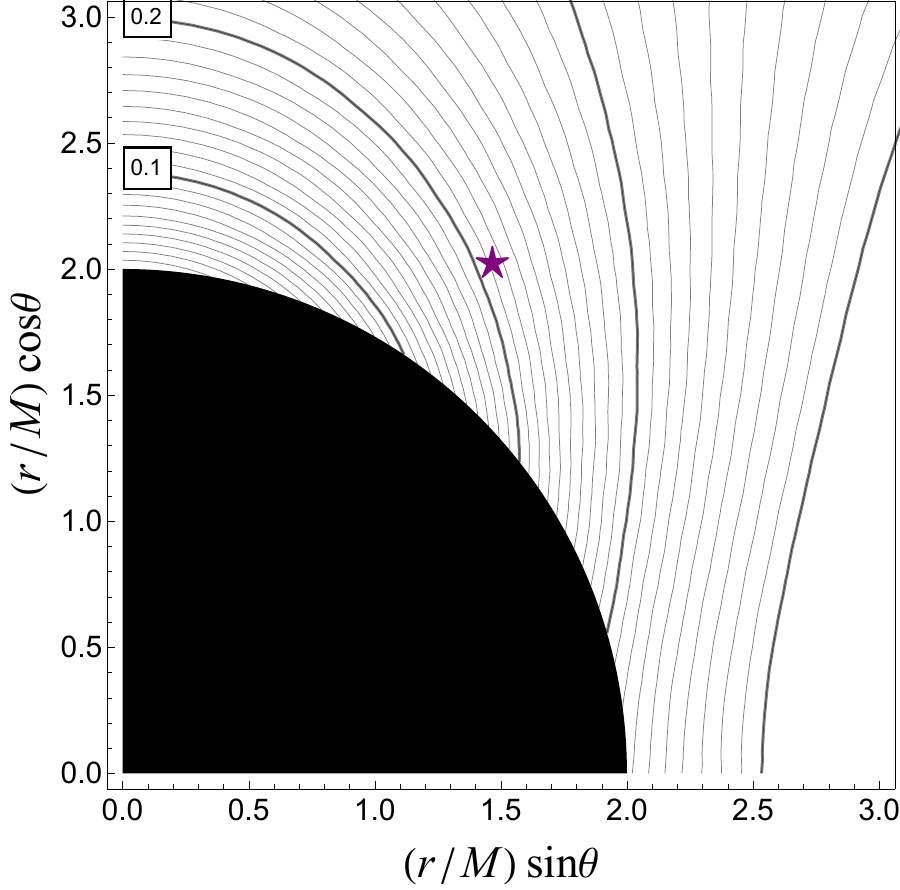}%
  }

  \subfloat[$a/M=0.80$\quad(global veiw)\label{subfig:sca_pot_a08dis}]{%
  \includegraphics[width=0.35\columnwidth]{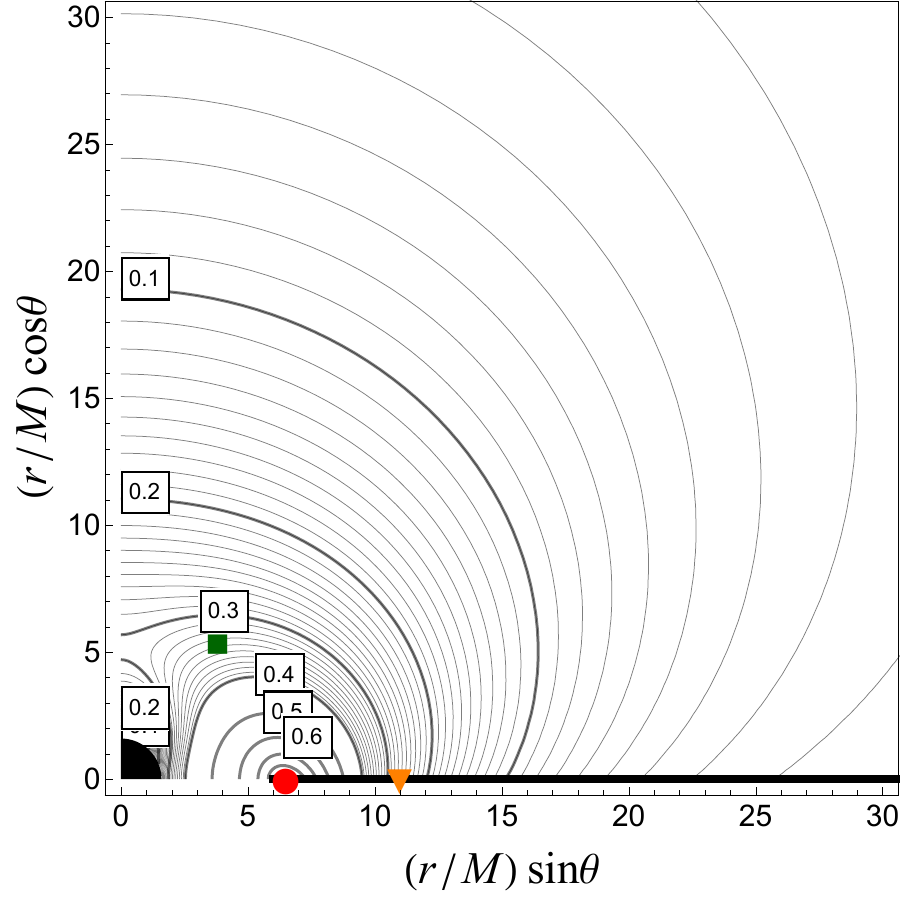}%
  }\quad\subfloat[$a/M=0.80$\quad(near horizon)\label{subfig:sca_pot_a08}]{%
  \includegraphics[width=0.35\columnwidth]{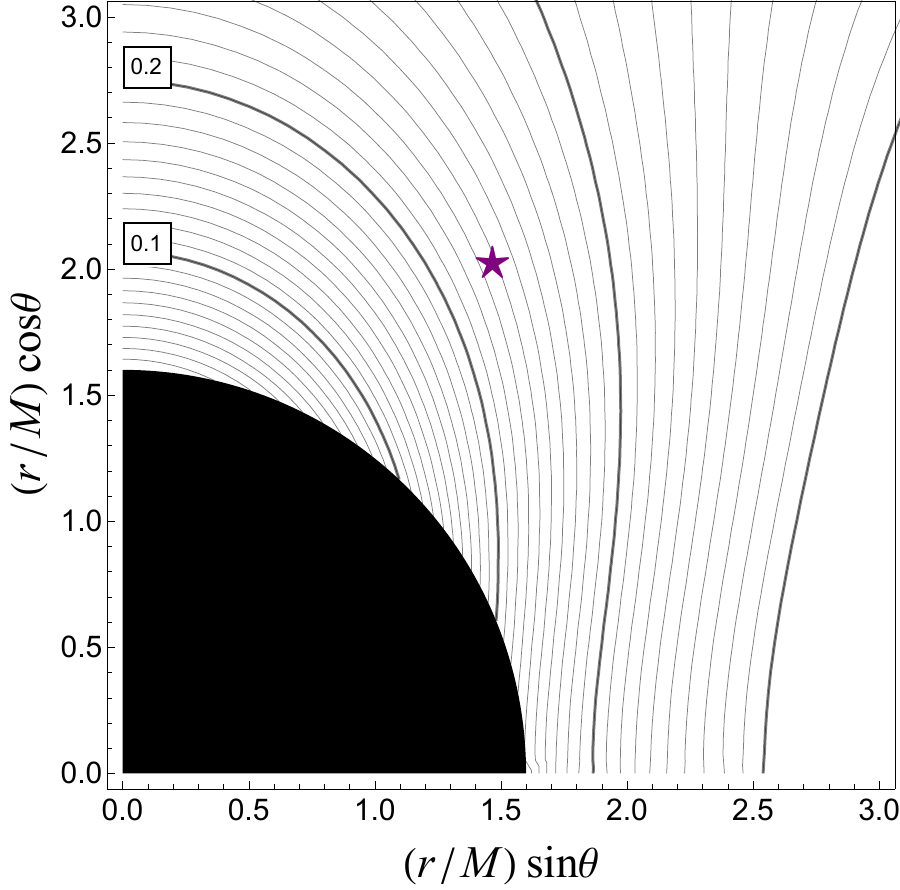}%
  }

  \subfloat[$a/M=0.9999$\quad(global veiw)\label{subfig:sca_pot_a09d4dis}]{%
  \includegraphics[width=0.35\columnwidth]{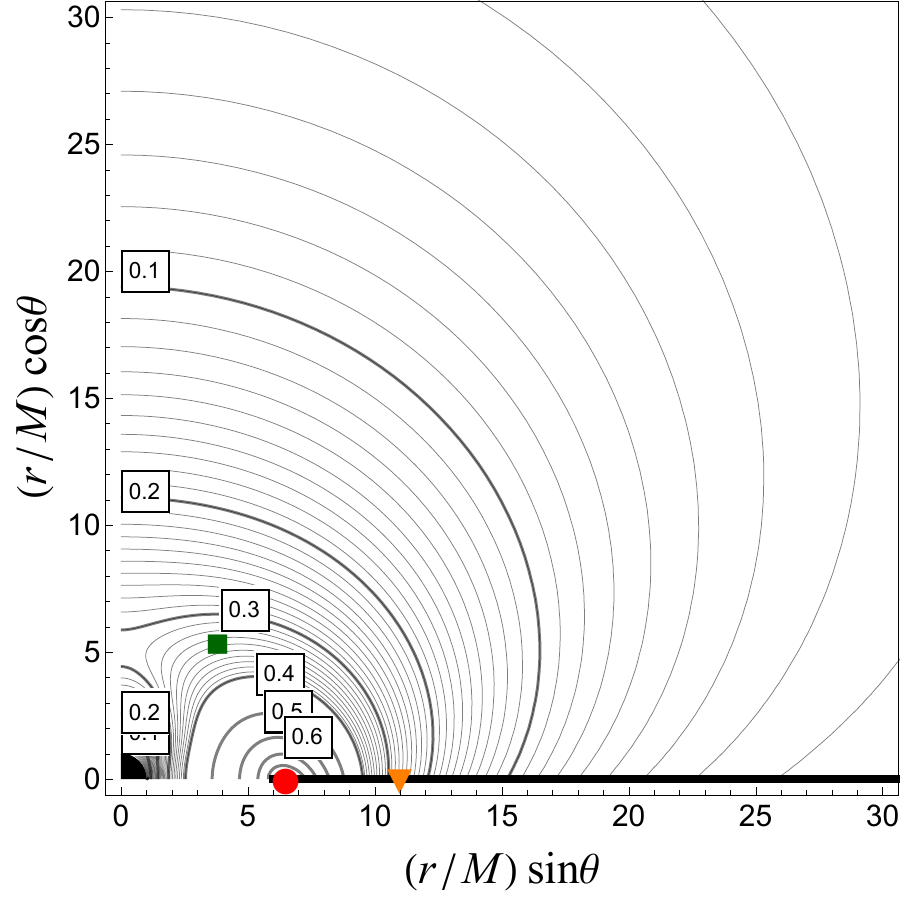}}\quad
  \subfloat[$a/M=0.9999$\quad(near horizon)\label{subfig:sca_pot_a09d4}]{%
  \includegraphics[width=0.35\columnwidth]{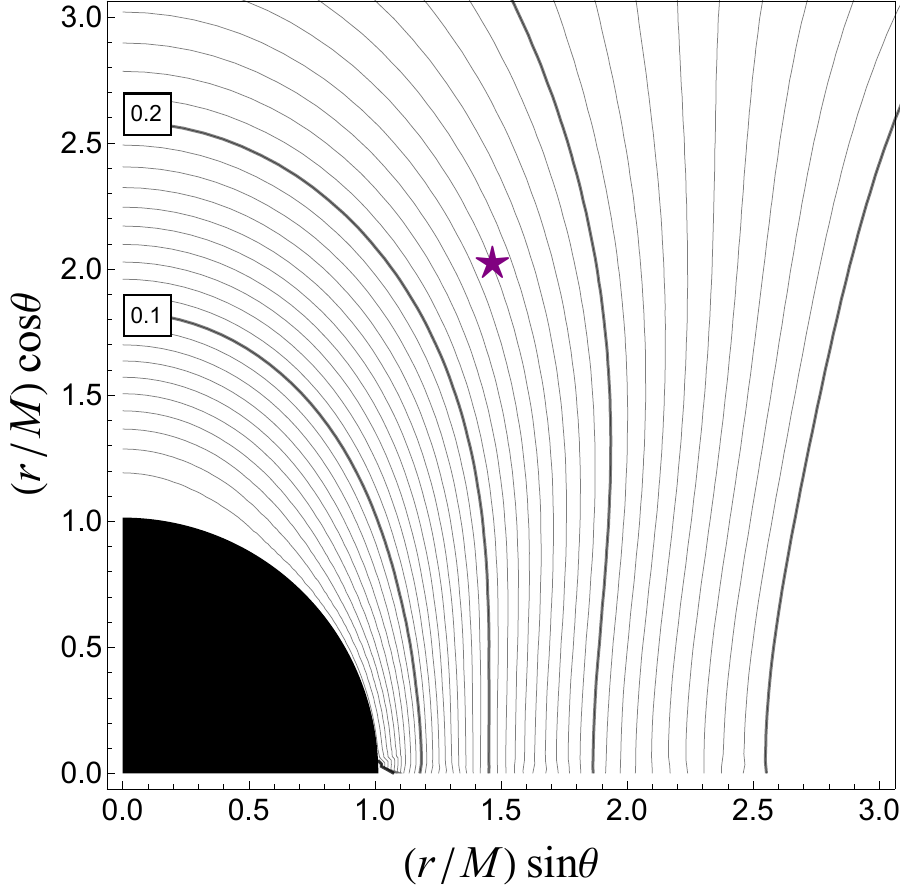}}
  \caption{
The contours of the electric potential $\Phi/\Phi^\mathrm{max}$ are shown for $a/M=0.0001,0.8,0.9999$. 
The global views are shown in the left panels \protect\subref{subfig:sca_pot_a00dis},\protect\subref{subfig:sca_pot_a08dis},\protect\subref{subfig:sca_pot_a09d4dis}, and the close-up views 
near the horizon are in the right panels \protect\subref{subfig:sca_pot_a00},\protect\subref{subfig:sca_pot_a08},\protect\subref{subfig:sca_pot_a09d4}. The quantity $\Phi^\mathrm{max}$ is proportional to the spin $a$ (see \figref{fig:dep_a:At}).
}\label{fig:scl_pot}
\end{figure}

\begin{figure}[H]
  \centering\includegraphics[width=0.4\columnwidth]{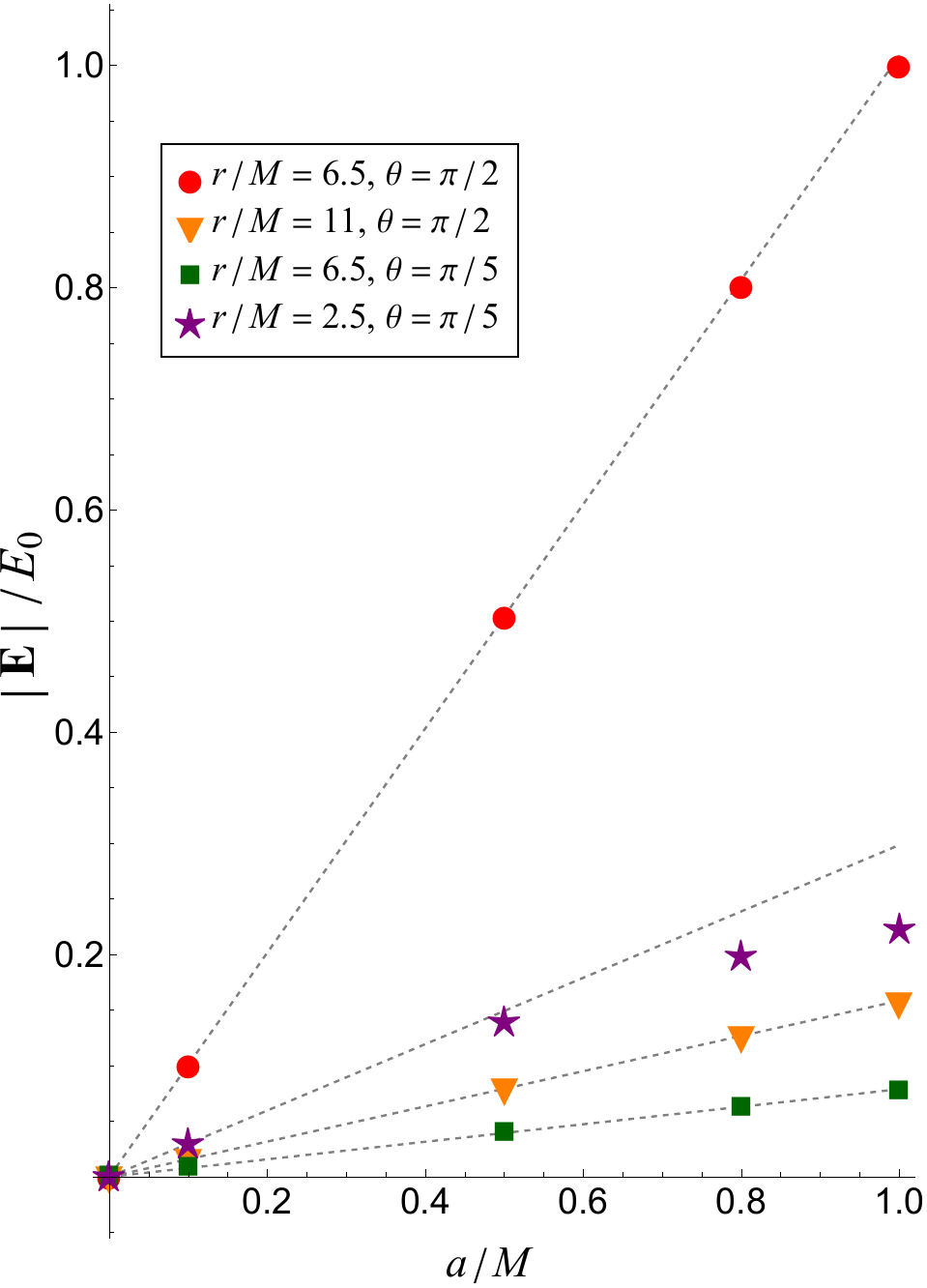}
  \caption{%
  The spin dependence of $|\bm{E}|$ at several points is shown in \figref{fig:scl_pot}. The red dot, orange triangle and green square correspond to points shown in the left column of \figref{fig:scl_pot}, while the purple star corresponds to a point shown in the right column of \figref{fig:scl_pot}. The amplitude $|\bm{E}|$ is normalized by $E_0$, where $E_0$ is the norm of the electric field with $a/M=0.9999$ at the point $r/M=6.5$ on the disk. 
  }\label{fig:dep_a:At}
\end{figure}

\begin{figure}[H]
	\centering\subfloat[Electric fields $\bm{E}$ \label{subfig:scl_coE_a099}]{%
	  \includegraphics[width=0.4\columnwidth]{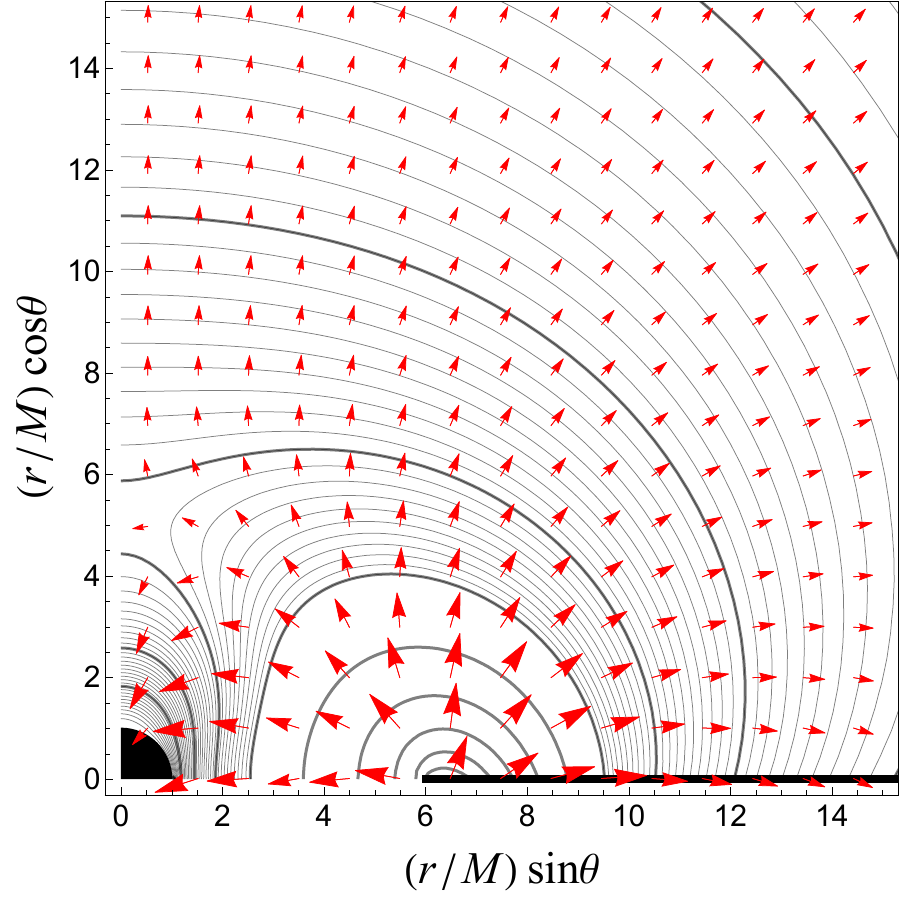}%
	  }\quad
	\subfloat[Electric flux density fields $\bm{D}$\label{subfig:coD_a099}]{%
	  \includegraphics[width=0.4\columnwidth]{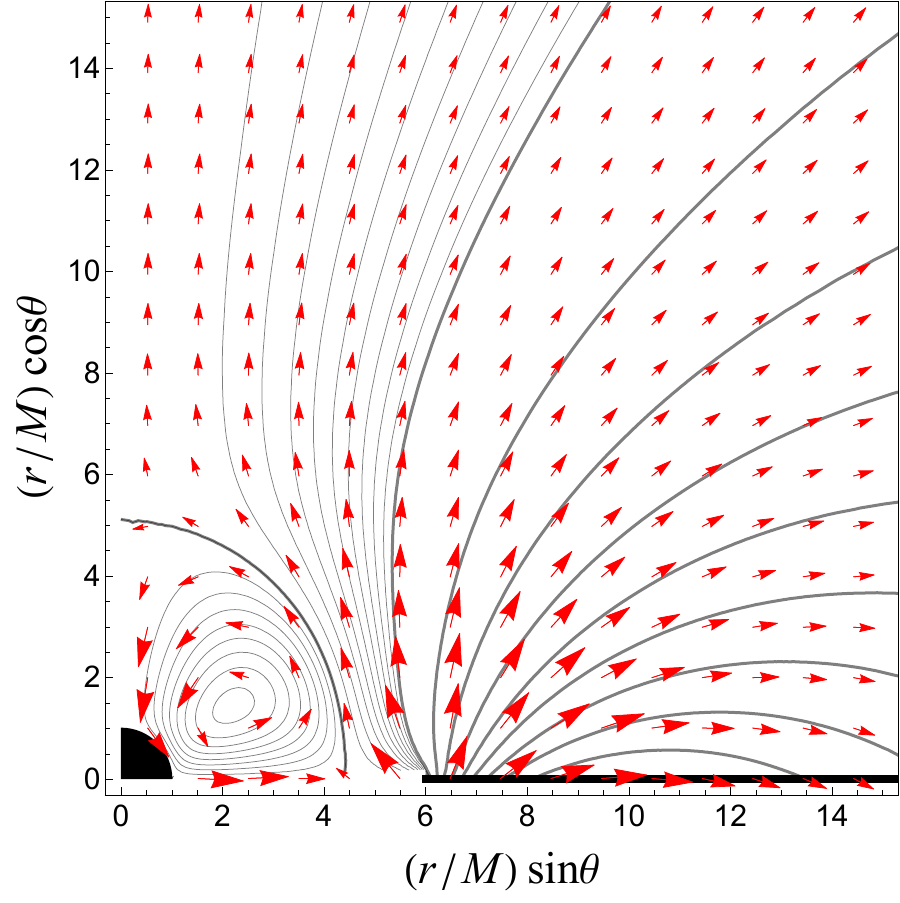}%
	  } 
  \caption{%
  The left panel \protect\subref{subfig:scl_coE_a099} shows the electric field $\E$ (red arrows) and the contours of the electric potential $\Phi$ (black lines) around the Kerr black hole with $a/M=0.9999$. The right panel \protect\subref{subfig:coD_a099} shows the electric flux density $\D$ (red arrows) and the integral curves of $\D$ (black lines) around the same black hole. 
  Near the black hole, $\E$ and $\D$ are clearly different, while, in the distant region, they coincide.}%
  \label{fig:scl_D}
\end{figure}

As shown in \figref{fig:scl_D}, we see the difference of the electric field $\E$ and the electric flux density $\D$ in our model. 
By definition $\E$ is rotation-free, while $\D$ can have the rotation.  
The difference is significant near the horizon where the shift vector $\beta^\varphi \coloneqq g_{t\varphi}/g_{\varphi\varphi}$ becomes large. 
In far region or on the rotation axis, $\E\parallel \D$ because $\beta^\varphi$ is small in the regions. 

Taking the reflection symmetry of $\bm{D}$ with respect to the equatorial plane 
into account, as shown in \figref{fig:scl_D}\subref{subfig:coD_a099}, 
$\boldsymbol \nabla\cdot \bm{D} \neq0$ on the disk. 
This is consistent with the charge distribution $\rho_e \propto j^t$, defined in Appendix \ref{app:EM_in_ZAMO}, where $j^t$ is proportional to $j^\varphi$ \eqref{jphi-jt}, and $j^\varphi$ is shown in \figref{fig:crnt_diff_a}. 
If we regard the black hole as a conducting sphere and the event horizon as the surface of the conductor, the configuration of $\D$ shown in \figref{fig:scl_D}\subref{subfig:coD_a099} suggests quadrupole-like charge separation on the horizon \cite{Thorne:1986jf}.

\subsection{Magnetic field-aligned components of electric field}

\begin{figure}[H]
  \centering\includegraphics[height=0.5\columnwidth]{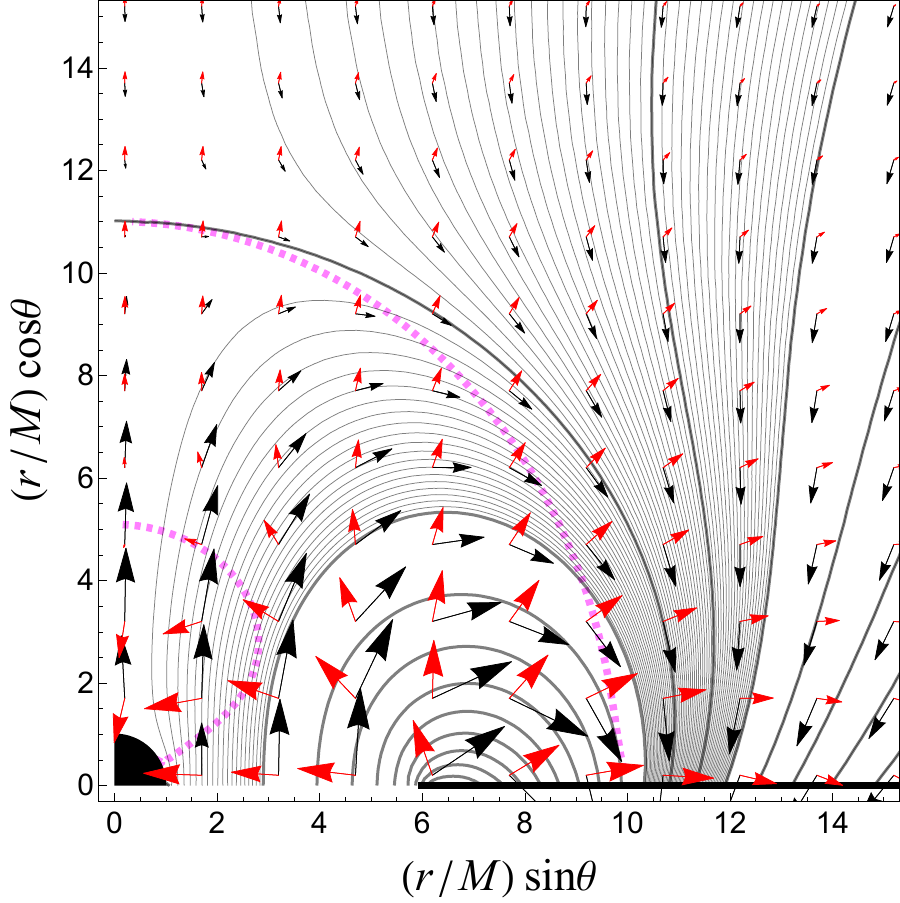}
  \caption{The vector plots of $\B$ (black arrows) and $\E$ (red arrows) are shown, overlaid on the contour plot of $\Psi$ for $a/M=0.9999$ in \figref{fig:mag_flx}\protect\subref{subfig:mag_flx_a09d4}. The vector fields $\bm{B}$ and $\bm{E}$ are orthogonal to each other on the dashed curves.}\label{fig:FsF}
\end{figure}

We study the direction of $\B\mbox{ and }\E$ in the vacuum magnetosphere in our model. 
As shown in \figref{fig:FsF}, we see that $\bm{B}\cdot\bm{E}\neq 0$ in almost all regions, and the directions of $\B$ and $\E$ are particularly parallel or anti-parallel near the rotation axis. 
Therefore, if we consider a charged particle whose motion is bounded on the magnetic flux line, we can expect that accelerations of charged particles by $\E$ occur along the magnetic flux line near the rotation axis. 
At the saddle point of $\Phi$ on the axis, the direction of acceleration changes from inward to outward or vice versa, depending on the sign of charge of particles.

In the present model, $F^{\mu\nu}F_{\mu\nu}= 2(|\boldsymbol{B}|^2-|\D|^2)$ is positive everywhere for all cases of $a$ except the small region around the zero point of $\B$. 
It means the magnetosphere is magnetically dominant almost everywhere. 
Indeed, the energy of the electric field is smaller than that of the magnetic flux density. 
However, the force by electric fields are important for the dynamics of charged particles. 
We estimate the ratio the electric force, $F_\mathrm{EF}$, to the gravity, $F_\mathrm{G}$, acting on charged particles. For example, consider SgrA*, whose mass is $4\times10^6M_\odot$, where $M_\odot$ is solar mass. 
The strength of the magnetic flux density in the central region is observed to be $|\bm{B}|\approx 10^1\sim 10^3$ Gauss \cite{Do:2019txfa,EventHorizonTelescope:2022urf}. 
From the relation between $|\bm{B}|$ near the inner edge and $|\bm{E}|$ near the separatrix on the rotation axis obtained numerically in our model, the ratio $F_\mathrm{EF}$ to $F_\mathrm{G}$ acting on protons (mass $m_\mathrm{p}$ and charge $e$), for example, is given by 
\begin{align}
    \frac{|F_\mathrm{EF}|}{|F_\mathrm{G}|} =  \frac{e|E|}{\frac{Gm_\mathrm{p} M}{r_\mathrm{sep}^2}}
    \simeq  10^6 \left(\frac{a}{M}\right)\left(\frac{M}{4\times10^6 M_\odot}\right)\left(\frac{|\bm{B}|}{10^2\ \mathrm{Gauss}}\right), \label{ratio:Flg_Fng}
\end{align}
where $F_\mathrm{G}$ is estimated by the Newtonian gravity, for simplicity. 
The relation \eqref{ratio:Flg_Fng} implies that the motion of the charged particles is driven by the electric force rather than by the gravity, and the particles may escape along the rotation axis.

\subsection{Charge accretion onto a black hole}\label{sec:acc}
In our model, we focus on the magnetic flux lines connecting the black hole and the disk. As shown in \figref{fig:mag_flx}, these flux lines have the footpoints located slightly inside the separatrix surface on the disk. 
Comparing \figref{fig:mag_flx} with \figref{fig:scl_pot}, we can see that there appear potential differences along these magnetic flux lines. 
For the rapidly rotating black hole case, the horizon becomes almost an equipotential surface, and \figref{fig:scl_pot} shows the potential difference between the footpoints of these magnetic flux lines satisfies $\Phi|_\mathrm{disk} - \Phi|_\mathrm{horizon}>0$. Therefore, it is expected that the positively charged particles fall into the black hole from the disk, and the accretion continues until the potential difference vanishes. 

In the following, for simplicity, we consider the magnetic flux lines connecting the event horizon near the north pole and the disk surface. 
We approximately evaluate the potential difference between the points $(r_\mathrm{H},0)$ and $(r_\textrm{sep},\pi/2)$, where the potential difference is maximal.
After the charging of the black hole with $Q_\mathrm{e}$, the accretion stops under the condition
\begin{equation}
     \left[\frac{r}{\Sigma}Q_\textrm{e}+\Phi\right]_{\substack{r=r_\mathrm{sep},\\ \theta=\pi/2\hphantom{,}}} - \left[\frac{r}{\Sigma}Q_\textrm{e}+\Phi\right]_{\substack{r=r_\mathrm{H},\\ \theta=0}}=0,\label{eq:eq_pot}
\end{equation}
where $rQ_\mathrm{e}/\Sigma$ term is the potential by the electric charge of the black hole. 
Hence, we have 
\begin{equation} 
     Q_\mathrm{e}  
     = \frac{2M r_\mathrm{sep}}{r_\mathrm{sep}-2M} 
     \Phi(r_\mathrm{sep}, \pi/2),
     \label{sol:acc_ch}
\end{equation}
because $\Phi(r_\textrm{H},0)=0$.

We estimate the amount of the accreted charge for SgrA*. 
From \eqref{sol:acc_ch} and the relation between $|\bm{B}|$ near the inner edge and $\Phi(r_\mathrm{sep}, \pi/2)$ obtained numerically in our model for a rapidly rotating black hole case, we have 
\begin{equation}
     Q_\textrm{e}\approx 1.8\times10^{15}\left(\frac{a}{M}\right)\left(\frac{M}{4\times10^6 M_\odot}\right)^2\left(\frac{|\bm{B}|}{10^2 \mathrm{Gauss}}\right)\textrm{C},\label{tot_ch}
\end{equation}
(see also \cite{Zajacek:2018ycb}). 
Therefore, the charge-to-mass ratio is obtained as 
\begin{equation}
     \frac{Q_\mathrm{e}}{M} 
     \approx 2.6\times10^{-12}\left(\frac{a}{M}\right)\left(\frac{M}{4\times10^6 M_\odot}\right)\left(\frac{|\bm{B}|}{10^2\mathrm{Gauss}}\right).
     \label{rel:ch_to_mss}
\end{equation}
Relation \eqref{rel:ch_to_mss} implies that the accreted charge-to-mass ratio onto the black hole is so small that the Kerr metric can be used. 
However, since the Coulomb force due to the accreted charge \eqref{tot_ch} is comparable to the electric force due to the black hole spin, the accreted charge on the black hole modifies the configuration of the electromagnetic field, and therefore the dynamics of charged particles would change. 
For slowly rotating black hole cases, because the horizon does not become an equipotential surface, the signs of the potential difference depend on the footpoints of the magnetic flux lines on the horizon. 
It would be hard to expect the charge accretion on the slowly rotating black hole.

\section{concluding remarks}\label{sec:dscss}
\subsection{Conclusions}

We have obtained stationary and axisymmetric solutions to the Maxwell equations using the Newman--Penrose formalism to construct a vacuum magnetosphere with a thin disk current in Kerr spacetimes. 
We matched the vacuum solutions in the northern and southern regions with the junction conditions given by the thin disk current on the equatorial plane. 
In our model, we assumed a toroidal current distribution on the disk that has an inner edge, and the 
direction of the currents is reversed at a radius that is several times the inner edge radius of the disk. 
In order to study how the spin parameter $a$ affects the electromagnetic field, we fixed the 
radius of the disk's inner edge and the current distribution on the disk.  

Hence, the magnetospheric solution has a separatrix surface of magnetic flux lines in a spherical shape. 
The location of the separatrix does not depend much on the spin parameter. 
In addition, an electric field is generated by the rotation of the Kerr black hole. 
The axisymmetry requires that both the magnetic and electric fields near the rotation axis are parallel to the axis. 
The presence of such an electric field component parallel to the magnetic field near the rotation axis implies the accelerations of charged particles along the magnetic field. 
Here, we found that a saddle point of the electric potential exists on the rotation axis inside the separatrix surface of the magnetic field. 
The presence of this saddle point on the axis means that the electric field switches its direction along the magnetic field lines.

The global structure of the magnetic field in the Kerr geometry is almost the same as the Schwarzschild case studied in TT01, while near the event horizon the structure of the electromagnetic field is similar to Wald's source free solution \cite{Wald:1974np}. 
This property suggests that, if a sourceless region surrounds the event horizon, the disk current 
does not affect electromagnetic fields in the neighborhood of the event horizon. 
In the extremal limit $a/M\rightarrow 1$, the black hole Meissner effect appears even in the presence of current on a disk with an inner edge, and the event horizon becomes an equipotential surface.

In our model, by the rotation of black hole, the charge distribution is associated with the toroidal current on the equatorial disk. 
Hence, the divergence of $\bm{D}$ appears on the disk from the reflection symmetry with respect to the equatorial plane. 
If we regard the black hole as a conducting sphere (see, e.g., \cite{Thorne:1986jf}), $\bm{D}$ on the event horizon indicates that a charge separation appears on the horizon. 
The charge distribution at the horizon is almost quadrupole, as well as the case of Wald's solution \cite{Wald:1974np}. 

Finally, we have considered charged particles falling into the black hole along the magnetic flux lines. 
In our model, there are magnetic flux lines connecting the disk and the horizon, while, in Wald's solution, there are only lines connecting the horizon and distant region.
In both cases, the potential difference appears along the magnetic flux lines. 
Here, in our model the charge accretes from the disk, while in Wald's case, the charge accretes from distant region. 
If we expect that the charging of the black hole stops when the potential difference vanishes, the value of the accreted charge in our model is estimated as almost the same as Wald's case \cite{Wald:1974np,Zajacek:2018ycb}. 
Such a charged black hole would generate large electric fields in the magnetosphere, and affect the motion of charged particles \cite{Zajacek:2018ycb,Nakao:2024weka}. 
In this way, the amount of charge on the black hole is extremely important. 

In our model, there exist the separatrix surface of $\Psi$ and the saddle point of $\Phi$, which depend on current distributions. We have only considered the model in which the total current vanishes. However, in models where the toroidal disk current does not reverse, neither a separatrix of the magnetic field nor a saddle point of the electric field would appear. 
Therefore, we should consider the more general case. In order to clarify an actual current distribution in a thin disk, a separate study of the internal structure of the thin disk should be required. 
This is a topic for future research. 

\subsection{Toward a force-free magnetosphere}
In this paper, we have studied a vacuum magnetosphere with a thin disk current.  
In order to consider the activity in high-energy astrophysical phenomena, we should consider plasma around the black hole. 
If a sufficient amount of charged particles enter the vacuum magnetosphere from the disk surface, the force-free magnetosphere would be achieved by shielding the electric field component along the magnetic field lines, that is, the charged particles are distributed so that the magnetic and electric fields become orthogonal to each other. 
Although a solution of the magnetosphere including plasmas has a different structure from the vacuum solution, the vacuum solution in some regions would be a good approximation of the force-free magnetosphere. 
In the force-free magnetosphere, the toroidal magnetic field produced by the electric current streaming along poloidal magnetic field lines should be considered. 
Here, the poloidal component of magnetic fields are self-consistently modified. 
Hence, it is important to investigate deviations of the poloidal magnetic field from the vacuum magnetosphere to consider high-energy astrophysical phenomena. 
Such an extension to the force-free magnetosphere is our future work.

\begin{acknowledgments}
We would like to thank K. -i. Nakao and H. Yoshino, in the astrophysics and gravity group in Osaka Metropolitan University, for valuable discussions. This work was partly supported by Osaka Central Advanced Mathematical Institute: MEXT Joint Usage/Research Center on Mathematics and Theoretical Physics Grant No. JPMXP0723833165. 
Y.E. was supported by JST SPRING, Grant No. JPMJSP2139.
\end{acknowledgments}

\pagebreak
  \makeatletter
    \renewcommand{\theequation}{%
    \Alph{section}.\arabic{equation}}
    \@addtoreset{equation}{section}
  \makeatother

\appendix

\section{Brief review of Wald's solution}\label{app:wald}
We consider a spacetime that admits a Killing vector field, say $\xi$. 
We take the gauge vector potential as 
\begin{align}
	A_\mu=\xi_\mu. 
\label{Wald_sol}
\end{align} 
Using the Killing equation $\nabla_\mu\xi_\nu+\nabla_\nu\xi_\mu =0$, 
we can show that the field strength $ F_{\mu\nu} :=\partial_\mu A_\nu-\partial_\nu A_\mu$ 
satisfies 
\begin{align}
  \nabla_\mu {F^\mu}_\nu 
	=&-2R_{\nu\mu}A^\mu. 
\end{align}
Therefore, if the spacetime is Ricci-flat, $A_\mu$ in \eqref{Wald_sol} is a solution that has the same symmetry 
as the spacetime to the vacuum Maxwell equation \cite{Wald:1974np}.  
The solution is called Wald's solution. 

Kerr black holes described by \eqref{met:Kerr} are Ricci flat spacetimes that admit two Killing vectors 
$\partial_t$ and $\partial_\varphi$. 
Therefore, $\bm{A} = c_1 \partial_t + c_2 \partial_\varphi$ 
is a stationary and axisymmetric solution to the vacuum Maxwell equation, 
where $c_1$ and $c_2$ are constants. 
If we require that the black hole has no electric charge, 
the constants should be chosen as
\begin{align}
  \bm{A} = a \partial_t +\frac{1}{2} \partial_\varphi, 
\label{wald_A}
\end{align}
where $a$ is the spin parameter of the Kerr black hole. 
It gives a homogeneous magnetic field at spatial infinity. 

The potential \eqref{wald_A} gives the magnetic flux function, $\Psi$, 
and electric potential, $\Phi$, defined by \eqref{mg_flx1} and \eqref{scl_ptn1} in the text, as 
\begin{align}
  \Psi =& A_\varphi 
	=-\frac{2Mr}{\Sigma}a^2\sin^2\theta 
		+ \frac{1}{2}\left(r^2+a^2+\frac{2Mr}{\Sigma}a^2\sin^2\theta\right)\sin^2\theta, 
\label{Psi_Wald}
\\
  \Phi =& -(A_t + a) 
	=- a\frac{2Mr}{\Sigma} +a \frac{Mr}{\Sigma}\sin^2\theta.
\label{Phi_Wald}
\end{align}
The electric potential appears if the spin parameter $a$ is non-vanishing. 
This implies that the electric field is generated by the rotation of black hole in the vacuum magnetosphere (see e.g. Fig.1 in \cite{Komissarov:2021vks}). 
Note that, from \eqref{Psi_Wald}, $\Psi \to 0$ for the extremely rotating limit, $a\to M$, i.e., all magnetic flux lines are blocked out of the horizon. 
This extreme case is just an analogy for the Meissner effect, and is called ``black hole Meissner effect'' \cite{Bicak:1985af,Komissarov:2007rc}. At the same time, we see $\Phi\to -M$ in \eqref{Phi_Wald} in the limit $a\to M$. 

\section{Electromagnetic fields in a stationary spacetime}\label{app:EM_in_ZAMO}
\subsection{\texorpdfstring{$(3+1)$}{3+1} decomposition of a statinary metric}
We assumed that spacetime is stationary, namely there exists the timelike Killing vector $\bm{\xi}$, which satisfies $\mathcal{L}_{\bm{\xi}} g_{\mu\nu}= 0$,
where $\mathcal{L}_{\boldsymbol{\xi}}$ is the Lie derivative along $\bm{\xi}$. 
The time coordinate is taken as 
\begin{equation}
  \bm{\xi} = \partial_t. \label{app:def_killing}
\end{equation}
Then, the metric is given in $(3+1)$ form, known as the ADM formalism, as  
\begin{equation}
  ds^2 = - \alpha^2dt^2 + \gamma_{ij}(dx^i+\beta^i dt) (dx^j+\beta^j dt)\label{app:mtr}
\end{equation}
where  $\alpha$ is called the ``lapse function", $\beta^i$ is the ``shift vector'', and $\gamma_{ij}$ is the three-dimensional metric tensor  on the $t=$constant spacelike hypersurfaces $\Sigma$. 

The 4-velocity of the local fiducial observer (FIDO), which is the unit normal to the spacelike hypersurface $\Sigma$, is 
\begin{equation}
  \bm{u} = -\alpha dt. \label{app:ZAMOf}
\end{equation}
The spatial metric $\gamma_{ij}$ coincides with the spatial components of the projection tensor
\begin{equation}
  \gamma_{\mu\nu} = g_{\mu\nu}+u_\mu u_\nu.
\end{equation}

Then, for arbitrary 4-vectors $M^\mu,\ N^\mu$, the inner product, the outer product, and  the norm on the hypersurface are defined as 
\begin{gather}
  \bm{M} \cdot \bm{N} \coloneqq \gamma_{\mu\nu} M^\mu N^\nu,\quad \left(\bm{M}\times \bm{N} \right)^\mu\coloneqq -\epsilon^{\mu\nu\alpha\beta} u_\nu M_\alpha N_\beta,\quad |\bm{M}| = \sqrt{\bm{M}\cdot\bm{M}},
\end{gather}
and the divergence and the rotation on the hypersurface are also defined as 
\begin{gather}
  \nabla \cdot \bm{M} \coloneqq \gamma_\mu^\nu\nabla_\nu (\gamma_\lambda^\mu M^\lambda),\quad \left(\nabla\times \bm{M} \right)^\mu\coloneqq -\epsilon^{\mu\nu\alpha\beta} u_\nu \gamma_\alpha^\lambda \gamma_\beta^\iota\nabla_\lambda (\gamma_\iota^\sigma M_\sigma),
\end{gather}
respectively, where $\epsilon_{\alpha\beta\gamma\delta}$ is the Levi-Civita tensor. 

\subsection{Electromagnetic fields in \texorpdfstring{$3+1$}{3+1} form in a stationary spacetime}

Consider an electromagnetic field in a stationary spacetime. In accordance with the $3 + 1$ decomposition
of the metric \eqref{app:mtr}, one can also make a similar decomposition for the components of the electromagnetic field. Following \cite{Komissarov:2004ms,Komissarov:2011vq}, we define 
\begin{align}
  &D^\mu \coloneqq \ F^{\mu\nu}u_\nu,\quad B^\mu \coloneqq -~ {}^\ast \!F^{\mu\nu}u_\nu,\quad \rho_e \coloneqq -j^\mu u_\mu,\label{app:vct_FIDO}\\
  &E_\mu \coloneqq \ F_{\mu\nu}\xi^\nu,\quad H_\mu \coloneqq -~ {}^\ast \!F_{\mu\nu}\xi^\nu, \quad J^\mu \coloneqq  \alpha j^\nu {\gamma_\nu}^\mu- \rho_e\beta^\mu,\label{app:vct_xi}
\end{align}
where $u_\mu$ and $\xi^\mu$ are introduced by \eqref{app:ZAMOf} and \eqref{app:def_killing}, $j^\mu$ is the four-vector of the electric current, $F_{\mu\nu}$ is the electromagnetic field strength, and $~ {}^\ast \!F^{\mu\nu}$ is the Hodge dual $~ {}^\ast \!F^{\mu\nu} = \frac{1}{2}\epsilon^{\mu\nu\alpha\beta} F_{\alpha\beta}$. In these definitions, the Maxwell equation by using the electric and magnetic flux density, $\bm{D}$, $\bm{B}$, defined by $\bm{u}$, and 
the electric and magnetic field, $\bm{E}$, $\bm{H}$, defined by $\bm{\xi}$ in a gravitational field can be reduced to a set of equations that are analogous to the Maxwell equations in flat spacetime
\footnote{Thorne \& MacDonald\cite{Thorne:1982we,Thorne:1982fn} defined spatial vectors of the electric and magnetic fields as measured by FIDOs, $\bm{u}$, and showed that the Maxwell equation can be reduced to the equations for two spatial vectors. The reduced equations derived by them are so complex as to involve the expansion and the shear of congruences of $\bm{u}$, explicitly.}. 
That is, the Maxwell equations are rewritten as
\begin{align}
  \nabla \cdot \bm{D} = 4\pi \rho_e,\qquad&\nabla \cdot \bm{B} = 0,\label{app:eq_Max1}\\
  \frac{\partial\bm{D}}{\partial t} - \nabla\times \bm{H} = - 4\pi\bm{J},\qquad&\frac{\partial\bm{B}}{\partial t} + \nabla \times \bm{E} = 0,\label{app:eq_Max2}
\end{align}
It should be noted that these four vectors $\bm{E,B,D,H}$ are related as 
\begin{align}
  \bm{E} = &\alpha \bm{D}+\bm{\beta}\times\bm{B},\label{app:ele_fld_flx}\\
  \bm{H} = &\alpha \bm{B}-\bm{\beta}\times\bm{D}.\label{app:emag_fld_flx}
\end{align}

\section{Gap of sources in the interval $r_\mathrm{H}<r<r_0$}\label{app:gap}
In Sec.\ref{sec:bc}, in addition to $Q = \tilde{Q} = 0$ and $\alpha^\Re_k=\tilde{\alpha}^\Re_k=0$, we have assumed $\mathcal{I}(u) = 0$ in the range $r_\mathrm{H}\leq r<r_0$ on the equatorial plane. This setting of  $\mathcal{I}(u)$ provides $j^\varphi = 0$ and $j^t=0$ in the interval $r_\mathrm{H} \leq r < r_0$ as shown below. 

At the horizon on the equatorial plane, from the solution \eqref{sol:phi1} with \eqref{Sl} and the junction condition \eqref{junction_current}, we see
\begin{align}
  \left.2\pi r^4 j^\varphi\right|_{r=r_H,\theta =\frac{\pi}{2}} 
  &= r_\mathrm{H}\int^\infty_0 dk \alpha_k^\Im (-\nu(\nu+1)) P_\nu(u=1)P_\nu(x=0) 
  \cr
  &= r_\mathrm{H} ~{\cal I}(u=1)=0,\label{app:crrt_I}
\end{align}
where in the second equality we have used the Maxwell equation \eqref{def_I}. 
Furthermore, from the definition of $\mathcal{I}(u)$ \eqref{def_I} and $d\mathcal{I}/dr = 0$ in the interval, it holds that 
\begin{align}
  \left.\frac{d}{dr}r^4j^\varphi\right|_{\theta =\frac{\pi}{2}} =0 \label{app:der_crrt_I}
\end{align} 
in the interval. From \eqref{app:crrt_I} and \eqref{app:der_crrt_I}, $j^\varphi$ vanishes in the interval $r_\mathrm{H} \leq r < r_0$. Therefore, by using the relation $j^t = a j^\varphi$, we confirm that $j^t$ vanishes on the equatorial plane in the interval.

\section{Meaning of assumptions in our model}\label{app:mn_srcs}
Our solution is characterized by $\alpha_k$ and $Q$. 
In order to investigate the meaning of $Q$, we calculate the general form of the toroidal current and the electric and magnetic charge. 

We define the toroidal current $I(r)$ as the integration of $J^\varphi$ over the poloidal surface intersecting the equatorial plane from the horizon $r_\mathrm{H}$ to $r$ on a $t=$ constant slice. From \eqref{eq:junction_j}, $I(r)$ is given by 
\begin{align}
  I(r) \coloneqq & \int_{\mathcal{S}_\varphi} d\boldsymbol{S}\cdot \boldsymbol{J}\notag\\
  =& -\frac{1}{4\pi}\int^r_{r_\mathrm{H}} d\bar{r} \left[\sqrt{-g}F^{\theta\varphi}\right]^+_-\notag\\
  = &\frac{1}{4\pi}
  \left[-\frac{1}{r}(Q^\Im  - \tilde{Q}^\Im ) + \frac{a}{r}\int^\infty_0  dk (\alpha_k^\Re+\tilde{\alpha}_k^\Re )P_\nu (u)\frac{dP_\nu (x=0)}{d\theta}-\frac{1}{r}\int^\infty_0  dk (\alpha_k^\Im - \tilde{\alpha}_k^\Im ) \Delta\frac{dP_\nu }{dr}P_\nu (x=0)\right]\notag\\
  &\quad - \frac{1}{4\pi}\left[-\frac{1}{r_\mathrm{H}}(Q^\Im  - \tilde{Q}^\Im ) + \frac{a}{r_\mathrm{H}}\int^\infty_0  dk (\alpha_k^\Re+\tilde{\alpha}_k^\Re )P_\nu (u=1)\frac{dP_\nu (x=0)}{d\theta}\right].\label{app:tot_ele_crrt}
\end{align}
where $\mathcal{S}_\varphi$ is the 2-dimensional surface with $\varphi=$constant between $\theta=\pi/2 - \epsilon$ and $\theta=\pi/2 +\epsilon$, and $F^{\mu\nu}$ is given by \eqref{Ftr} - \eqref{B_phi} with \eqref{sol:phi2}, \eqref{sol:phi0} and \eqref{sol:phi1}. 

From the Gauss law \eqref{app:eq_Max1}, the general form of the electric charge $\mathcal{Q}_\mathrm{e} (r)$ and the magnetic charge $\mathcal{Q}_\mathrm{m} (r)$ involved within $r=$ constant surface are defined by 
\begin{align}
  \mathcal{Q}_\mathrm{e} (r) \coloneqq & \int_\mathcal{V} dV \rho_\mathrm{e}
  = \frac{1}{4\pi}\int^{2\pi}_0 d\varphi\int^\pi_0 d\theta\sqrt{-g} F^{tr}\cr
  = &\frac{1}{2} (Q^\Re +\tilde{Q}^\Re ) + \frac{1}{2}\int^\infty_0  dk (\alpha^\Re_k+\tilde{ \alpha}^\Re_k )\left[\frac{r^2+a^2}{r}P_\nu (u) - \frac{1}{\lambda}\Delta\frac{dP_\nu }{dr}\right]\frac{dP_\nu (x=0)}{d\theta} \notag\\
  &-\frac{a}{2r}(Q^\Im -\tilde{Q}^\Im ) - \frac{a}{2r}\int^\infty_0  dk (\alpha^\Im_k-\tilde{\alpha}^\Im_k)  \Delta \frac{dP_\nu }{dr}P_\nu (x=0),\label{app:tot_ele_ch}\\
  \mathcal{Q}_\mathrm{m} (r) \coloneqq & \frac{1}{4\pi}\int^{2\pi}_0 d\varphi\int^\pi_0 d\theta \sqrt{-g}~ {}^\ast \!F^{rt} = \frac{1}{4\pi}\int^{2\pi}_0 d\varphi\int^\pi_0 d\theta F_{\theta\varphi}\notag\\
  = & \frac{a}{2r}(Q^\Re -\tilde{Q}^\Re ) + \frac{a}{2r}\int^\infty_0  dk (\alpha^\Re_k -\tilde{\alpha}^\Re_k)\Delta \frac{dP_\nu }{dr}P_\nu (x=0)\notag\\
  & + \frac{1}{2}(Q^\Im +\tilde{Q}^\Im ) + \frac{1}{2}\int^\infty_0  dk (\alpha^\Im_k +\tilde{\alpha}^\Im_k )\left[\frac{r^2+a^2}{r}P_\nu (u)-\frac{1}{\lambda}\Delta\frac{dP_\nu }{dr}\right]\frac{dP_\nu (x=0)}{d\theta},\label{app:tot_mag_ch}
\end{align}
respectively, where $\mathcal{V}$ is the volume within the surface defined by the $r=$ constant on a $t=$ constant slice. 

When, for simplicity, $\alpha_k = \tilde{\alpha}_k=0$ and $Q^\Re =\tilde{Q}^\Re  \neq 0$, $Q^\Im=\tilde{Q}^\Im \neq 0$
is assumed, it is obvious that the integration constant $Q^\Re $ and $Q^\Im$ mean the electric and magnetic charge of the black hole, respectively, since $\mathcal{Q}_\mathrm{e}(r_\mathrm{H}) = Q^\Re $ and $\mathcal{Q}_\mathrm{m}(r_\mathrm{H}) = Q^\Im$. 

In our model, we have assumed no electric or magnetic charge of the black hole, i.e., $Q =\tilde{Q} = 0$.
Therefore, $\mathcal{Q}_\mathrm{m}(r)$ given by \eqref{app:tot_mag_ch} vanishes by using the relation $\tilde{\alpha}_k=\alpha_k^*$ in \eqref{alpha_symm}, that is, there is no magnetic charge anywhere. 
Furthermore, we have assumed $\alpha^\Re_k=\tilde{\alpha}^\Re_k=0$, which means no disk charge in the limit $a\to 0$. It is shown from \eqref{app:tot_ele_crrt} and \eqref{app:tot_ele_ch} that $I^\mathrm{tot}=I(\infty)$ and $\mathcal{Q}_\mathrm{e}^\mathrm{tot}=\mathcal{Q}_\mathrm{e} (\infty)$ vanish. Thus, there are no total toroidal current and no total electric charge on the disk regardless of the choice of $\alpha^\Im_k$.

\newpage

\bibliography{BH_mag,text,observation}

\end{document}